\documentclass[aps,preprint, groupedaddress]{revtex4-2}

\usepackage{graphicx}
\usepackage{multirow}
\usepackage{subfig}
\usepackage{amsmath,amsfonts,amssymb}
\usepackage{bbold, bm}
\usepackage{subfig}
\DeclareMathOperator{\argmin}{argmin}
\usepackage{xcolor}
\usepackage{float}
\begin{document}


\title{Evolutionary computing and machine learning for the discovering  of low-energy defect configurations}


\author{Marco Arrigoni}
\email{marco.arrigoni@tuwien.ac.at}
\author{ Georg K. H. Madsen}
\affiliation{Institute of Materials Chemistry, TU Wien, A-1060 Vienna, Austria}

\date{\today}

\begin{abstract}
Density functional theory (DFT) has become a standard tool for the study of point defects in materials.
However, finding the most stable defective structures remains a very challenging task as it involves the solution of a multimodal optimization problem with a high-dimensional objective function.
Hitherto, the approaches most commonly used to tackle this problem have been mostly empirical, heuristic and/or based on domain knowledge. 
In this contribution, we describe an approach for  exploring the potential energy surface based on the covariance matrix adaption evolution strategy (CMA-ES) and supervised and unsupervised machine learning models. We show how the original CMA-ES can be modified to suit the specific problem of DFT studies of point defects in the dilute limit. The resulting algorithm depends only on a limited set of physically interpretable hyperparameters. The approach offers a robust and systematic way for finding low-energy configurations of point defects in solids. We demonstrate the applicability and moderate computational cost on the intrinsic defects in silicon. We also apply the methodology to the neutral oxygen vacancy oxygen vacancy in TiO$_2$ anatase and reproduce the known defect structures. Furthermore, a new defect structure, stable at the level of hybrid density functional theory and characterized by a delocalized electronic structure, is found for this system.
 
\end{abstract}


\maketitle

\section{Introduction}
Point defects are ubiquitous in materials and their presence is able to modify the system behavior considerably. The controlled inclusion of defects in semiconductor and insulator materials can be employed to tune electronic, optoelectronic, electrochemical and catalytical properties and is commonly exploited in technological applications. On the other hand, the presence of point defects is not always beneficial, but can also lead to a degradation of the performance and lifetime of a device \cite{book:Pizzini,Queisser-1998,book:McCluskey, Laks-1991,Zunger-2003,Maier-2005,Alamo-2011,Yu-2016,Walsh-2017}.
The desire to control the properties of defective systems requires a detailed understanding of the interaction of point defects with the host material, which has spurred the development of specific characterization methods.

First-principles calculations, based mostly on density functional theory (DFT)\cite{Kohn-1965}, have become a standard tool in the study of point defects in solids. In addition to their predictive power, first-principles calculations are able to provide information on several properties of interest, such as the atomic-scale geometric structure, the electronic density and wavefunction and the defect thermochemistry, quantities that cannot be easily obtained from experiments \cite{Walle-2004,PDRev,DefectsSemiconductors}. 
As the defect concentrations most relevant for technological applications are generally very low, the first-principles study of point defects in solids has mostly been aimed at the so called \emph{dilute limit}, where defect-defect interactions are considered to be negligible. To perform such calculations, one commonly employs the supercell method, in which a sufficiently large portion of the host material is taken as the simulation cell and the point defect of interest is then introduced therein. In order to minimize the spurious defect-defect interactions arising from the application of periodic boundary conditions, relatively large supercells with around 100 atoms or more might be required.

The large simulation cells, combined with defect-induced effects that strongly alter the energy landscape, make the systematic search for low-energy minima of the potential energy surface (PES) a formidable task. The approach most commonly used to tackle this problem is based on domain knowledge: the most likely low-energy defect configurations are selected according to intuition and/or the results obtained on analogues systems. Once the defect is introduced in the chosen configuration, the atomic positions are relaxed to a minimum of the PES using gradient-based optimization algorithms. However, there is no guarantee that the found minimum corresponds to the most stable defect configuration. Furthermore, a more complete understanding of the low-energy minima in the PES is also sought after, as metastable defect configurations can be responsible for phenomena that cannot be attributed to the ground-state \cite{Lany-2005,Kundu_PRM19}. Searches based on previous knowledge are evidently unsatisfactory as the curse of dimensionality will inevitably make unstructured heuristic methods perform poorly.

Finding the ground-state and low-energy meta-stable structures of a point defect is a problem of multimodal optimization. Several algorithms have been proposed to tackle such problems with evolutionary algorithms (EAs) being amongst the most popular and successful ones. EAs 
 are a family of metaheuristic algorithms whose design is inspired by the Darwinian theory of natural evolution: a set of candidate solutions is generated and iteratively evolved through the application of genetic operators that  select, recombine and mutate the candidate solutions in order to generate new ones with higher fitness \cite{Eiben-2015, Omid-2017}. Although many components of EAs depend on stochastic processes and convergence to the global minimum is not guaranteed, their usefulness in optimization problems has been demonstrated in several complex problems. Genetic algorithms (GAs), the oldest, and perhaps most famous, class of EAs were adopted early on in computational materials science  \cite{Deaven-1995} and more recently also in combination with first-principles DFT studies \cite{Jacobsen-2002,Oganov-2006,Martinez-2011,Vilhelmsen-2012,Lysgaard-2014,Merte-2017}. 
There is however only a limited number of studies applying GAs or related methods to solids containing containing point defects \cite{Patra-2018,Atilgan-2018,Cheng-2020} and, to the best of out knowledge, no study has so far tackled the problem of finding the low-energy configurations of isolated point defects in solids. 

In this contribution, we present a computational approach that aims exactly at the solution of this problem. Our approach is based on the covariance matrix adaptation evolution strategy (CMA-ES) algorithm \cite{Hansen-2001}.  ESs  are mainly employed for solving continuous-parameter optimization tasks and, although the distinction between the two classes has become more blurred in recent years, the main difference between ESs and GAs is the presence of self-adapting endogenous strategy parameters in the former. Such strategy parameters determine  the distribution of the candidate solutions and are adapted during the evolutionary process according to the individuals' fitnesses.
ESs possess some advantages that make them a promising alternative to the more popular GAs. The flexibility of GAs is endowed by the existence of a wide variety of genetic operators. This might be a benefit for maximizing the performance of a specific algorithm for a specific problem but can also require a lot of fine-tuning, as each kind of genetic operator depends on some hyperparameters, whose optimal values generally have to be obtained heuristically. The choice of genetic operators in ESs is typically much more restricted facilitating the systematic optimization of the hyperparameters. In the present work we demonstrate how the original CMA-ES can be modified to suit DFT studies of point defects in the dilute limit, leading to an algorithm which depends on a limited set of physically interpretable hyperparameters.

Characteristic of all  EAs, is the generation of several structures before a converged solution is obtained making methods able to make use of this information particularly useful. We propose two methods based on machine learning (ML) to do this. First of all, in order to discover additional low-energy minima, we propose a method employing an unsupervised-learning post-processing step which is able to exploit the configurations explored by the EA and discover efficiently defective structures leading to low-energy minima which do not coincide with the solution found by the EA itself. Furthermore, we address one of the main drawbacks common to all EAs, namely the need of several evaluations of the fitness function before an optimal solution is found, which make their use in first-principles studies computationally very demanding. 
The recent development of 
accurate supervised ML methods \cite{Behler-2007,Bartok-2010,Rupp-2012,DeVita-2015,Bartok-2018} 
has allowed the use of ML metamodels  with the ability of approximating accurately the DFT PES in conjunction with EAs  \cite{Jennings-2019,Bisbo-2020}. We investigate the augmentation of our proposed EA by the inclusion of a ML metamodel based on Gaussian processes (GPs) in order to substantially reduce the computational burden.

The rest of the paper is organized as follows: section \ref{sec:Method} illustrates all the components of the proposed computational approach in detail and analyzes their contribution to its overall performance. In particular, subsection \ref{subsec:EA} is concerned with the description of the EA, subsection \ref{subsec:CLUS} describes the unsupervised post-processing step and subsection \ref{subsec:GPR} the ML regressor metamodel. The capabilities of the specific method outlined in each subsection is shown by considering intrinsic impurities in silicon as test cases.
Finally, in Section   \ref{sec:Results} the proposed approach is applied for the study of uncharged oxygen vacancy, $\square_\text{O}$, in TiO$_2$ anatase. We show that the method described in this work is able to find a low-energy configuration for the defect which is stable at the level of hybrid functionals and that has not been hitherto considered.

\section{Methodology \label{sec:Method} } 
\subsection{Evolutionary Algorithm \label{subsec:EA}}

The CMA-ES \cite{Hansen-2001,Hansen-2004,Hansen-2006,Hansen-2016} has been successfully applied in many complex problems ranging from engineering to artificial intelligence \cite{CMAES-site}. 
One of the main improvements of the CMA-ES over previous ESs was the use of evolution paths  \cite{Hansen-2001}, which contain, not  only information on a single generation, but on the whole evolutionary process and can be used to effectively adapt the strategy parameters. Other characteristics of the CMA-ES that make the use of this algorithm more advantageous and require less problem-dependent tuning are its invariance with respect to any strictly monotonic transformation of the fitness function, invariance with respect to translations and both proper and improper rotations of the coordinate system \cite{Hansen-2006}. 

In the following, we start from the implementation presented in reference \onlinecite{Hansen-2016} and adapt it in order to solve the particular optimization problem of interest for this work: finding the stable structures of an isolated point defect in a supercell. If not  stated otherwise, the hyperparameter values are set to the default ones proposed in that study.
The aim of the CMA-ES is to find the vector that minimizes a cost function  $f$: 
\begin{equation}
\label{eq:minim}
\mathbf{x}^\star = \underset{\mathbf{x} \in {\rm I\!R}^d}{\argmin}\,  f(\mathbf{x}).
\end{equation}
Here $\mathbf{x}$ is a vector in the genotype space, ${\rm I\!R}^d$, and  represents an individual. Considering the particular problem of interest in this work, we can consider an individual as a particular defect-containing supercell, whose cell parameters are fixed to those of the host material, but the atomic positions are free to vary. We chose the natural genetic representation of $\mathbf{x}$ being the vector of the atomic Cartesian coordinates, hence, $d = 3N$, with $N$ being the number of atoms in the supercell. Moreover, the natural choice for the cost function is the potential energy of the system.
In order to search the genotype space, the CMA-ES samples new individuals from a multivariate Gaussian distribution, which parameters are adapted during the evolutionary process. At generation $g$ a population made of $\lambda$ individuals, $\{\mathbf{x}^{(g)}_k\}_{k = 1, \dots, \lambda}$ is sampled as:
\begin{equation}
\label{eq:sampling}
\mathbf{x}^{(g)}_k \sim \text{N}\left(\mathbf{m}^{(g)}, (\sigma^{(g)})^2 \mathbf{C}^{(g)} \right).
\end{equation}
where $\mathbf{m}^{(g)} \in {\rm I\!R}^d$
is the mean of the normal distribution and $(\sigma^{(g)})^2 \mathbf{C}^{(g)}$ its covariance matrix. Specifically $\mathbf{C}^{(g)} \in {\rm I\!R}^{d\times d}$ is a symmetric and positive semidefinite matrix and $\sigma^{(g)} \in {\rm I\!R}$ a global step-size. All these quantities form the strategy parameters of the algorithm and their values are updated during the evolutionary process.

Once the fitness function has been calculated for each individual, the algorithm employs the truncation selection operator for selecting the parent individuals which will be used to generate the offspring of the next generation. In truncation selection, $\mu$ parents are selected deterministically according to their fitness. For this reason, it is convenient to rank all the individuals according to this property; in particular, with $\{ \mathbf{x}_{(k)}^{(g)} \}_{k=1, \dots, \lambda}$ we denote the individuals of generation $g$ rearranged in order of descending fitness. The population size $\lambda$ generally is chosen according to:
\begin{equation}
\lambda = 4 + \lfloor 3 \ln d \rfloor,
\end{equation}
as suggested in the original work \cite{Hansen-2001}. This quantity is hence increasing very slowly with the search-space dimension. This is particularly convenient for high-dimensional problems as it entails a reduced computational burden when the evaluation of the fitness function is expensive.  
 
For defect calculations, the first thing to consider is that relative large supercells are needed. The typical displacement field produced in a material by the defect-induced distortions, however, quickly subsides as the distance from the defect increases. 
This local nature of the structural perturbation suggests to turn the original optimization problem into a constrained one with a lower dimensionality.  Given that $\mathbf{x}(i)$ indicates the coordinates of atom $i$ in the supercell, one can thus introduce a hard cutoff radius $c_n$ such that only the atoms with $||\mathbf{x}(i) - \mathbf{x}_d||_2 < c_n$  contribute to the total number of degrees of freedom. Here $\mathbf{x}_d$ indicates the coordinates of the defect center and the distances are calculated taking the periodic boundary conditions into account. This has the advantage of reducing the dimensionality of the problem from $3N$ to $3s$, where $s$ is the number of atoms within the cutoff. The choice of the value $c_n$ is critical, as in the constrained optimization the algorithm will converge to the global minimum $\mathbf{x}^\ddagger \in {\rm I\!R}^d$ of the constrained PES. A too small value would impose too strong constraints, altering considerably the PES and ultimately making $\mathbf{x}^\ddagger$ substantially different from the true global minimum $\mathbf{x}^\star$, so that, even if the atomic positions of the solution configuration are relaxed without any constraint, the relaxed configuration does not coincide with $\mathbf{x}^\star$.

The localized nature of many point defects suggests another improvement on the algorithm. Normally, the matrix $\mathbf{C}$ is initialized to the identity matrix; however for the problem of interest, this initialization will cause the algorithm to initially sample highly disordered configurations with a very high energy. This is especially so because the initial global step size, $\sigma^{(0)}$, should be large enough in order to ensure that the PES is sampled thoroughly. To avoid this, we initialize $\mathbf{C}$ in such a way that displacements for the atoms close to the defective site are endowed with a larger variance. This permits an exhaustive sampling of the most relevant part of the PES and avoids the generation of high-energy structurees. Specifically, the covariance matrix is initialized as:
\begin{equation}
\label{eq:c_init}
\mathbf{C}^{(0)} = \mathbf{I}_{3s} + \sum_{j=1}^{3s} c_r^2(\pi_3(j)) \mathbf{e}_{j}\mathbf{e}_{j}^\intercal.
\end{equation}
The vectors of the standard basis of ${\rm I\!R}^{3s} $ are represented by $\mathbf{e}_{j}$. The function $\pi_3 : {\rm I\!N} \rightarrow {\rm I\!N}$ maps indexes running over components of $\mathbf{x}$ onto indexes running over the $s$ atoms within the cutoff according to:
\begin{equation}
\pi_3(n) = \left(n \, div \, 3 \right) + 1,
\end{equation}
where $div$ indicates the integer division. The coefficients $c_r(i)$ are given by:
\begin{equation}
\label{eq:c_r}
c_r(i) = \frac{c_r}{1 + ||\mathbf{x}^{(0)}(i) - \mathbf{x}_d||_2},  \qquad i = 1, \dots, s;
\end{equation}
 The addition of one (\AA) to the distance from the defect avoids the appearance of singularities when the defect is not a vacancy. The summation term on the right-hand-side of equation~\eqref{eq:c_init} is a matrix of rank $3s$ and $c_r$ is the hyperparameter defining the relative  importance of this matrix in the initialization of  $\mathbf{C}$.

The last modification is due to the fact that forces in first-principles calculations can be conveniently  obtained from the Hellmann-Feynman theorem. So, while the CMA-ES was initially proposed for black-box optimization problems, where one can only evaluate $f(\mathbf{x})$, the information given by $\nabla f(\mathbf{x})$ is generally available in DFT calculations, and  it can be used in the algorithm.
In particular, we update the the mean of the distribution according to:
\begin{equation}
\label{eq:mean_update_grad}
\mathbf{m}^{(g + 1)}  = \mathbf{m}^{(g)} + c_\text{m} \sum_{k=1}^\mu w_k \left[ \left(\mathbf{x}_{(k)}^{(g)} - \mathbf{m}^{(g)}\right) - c_\alpha^{(g + 1)} \nabla f(\mathbf{x}_{(k)} ^{(g)})\right],
\end{equation}
where $c_\text{m}$ is a learning rate which has been set to one in this work. The weigths $w_k > 0$ are characterized by a magnitude that is not increasing with $k$ and such that $\sum_{k=1}^\mu w_k = 1$.
In addition to the first term, which is also present in the original implementation of the CMA-ES, equation~\eqref{eq:mean_update_grad} also includes the term depending on the cost function gradient.  
 This update not only  forces the distribution mean to move more towards the basins where high-fitness individuals are located, but takes into account the basins' slope. The hyperparameter $c_\alpha$ determines the relative importance of the gradient term in updating this mean. Ideally, its value should be small far from the minimum; otherwise the atomic forces might be very large, which in turn would shift the distribution mean towards regions of the PES far from any minimum. As a consequence, individuals generated in the following generation would be characterized by very high energies and the convergence rate of the algorithm would be reduced considerably or fail completely. Conversely, in order to speed-up the convergence, one would want a large value of $c_\alpha$ close to the minimum and/or in flat regions of the PES where the cost function is almost constant. This suggests to adapt the value of $c_\alpha$ iteratively. In order to satisfies the former two conditions, we employed an update in the form:
\begin{equation}
\label{eq:c_alpha_update}
c_\alpha^{(g + 1)} = c_\alpha^{(g )} \exp\left( c_\beta \left[ 1 - \frac{\sigma^{(g)} ||\sum_{k=1}^\mu w_k  \nabla f(\mathbf{x}_{(k)} ^{(g)})||_2}{\text{E}[||Z||_2]}  \right] \right),
\end{equation}
where $Z \sim \text{N}(\mathbf{0}, \mathbf{I}_{3s})$ and $\text{E}$ indicates the expectation of a random variable. The hyperparameter $c_\beta$ has been set to $0.1 \frac{\sigma^{(0)}}{\text{eV}}$ throughout the present work.

While the algorithm is characterized by the presence of several hyperparameters, in practice the default values suggested in reference~\onlinecite{Hansen-2016} have been found to be adequate. 
This makes $\sigma^{(0)}$, $\mathbf{m}^{(0)}$ and $c_n, c_r $ and $c_\alpha^{(0)}$ the only problem-dependent quantities. This is a small set of  parameters, all of which have a straightforward physical interpretation, thus making their tuning an uncomplicated task. Moreover, it is  natural to initialize $\mathbf{m}^{(0)}$ with the coordinates of an intuitive initial defect configuration and to let $\mathbf{x}_k^{(0)} = \mathbf{m}^{(0)}$ for all the individuals of the initial generation. For example,  for a system with a vacancy, such initial configuration would consists in the supercell with the removed atom, before any optimization of the atomic positions takes place. Similarly, for an interstitial impurity, the initial configuration would be the unrelaxed supercell with the defect placed in some reasonable interstitial site. We call such an initial defect configuration  the \emph{population founder}.

\subsection{Performance: the Si$_i$ in Si with an analytic potential\label{subsec:Si_int}}
  To investigate the performance of the modified algorithm for the optimization problem of finding low-energy configurations of point defects in solids, we first considered the Si interstitial in bulk silicon using a supercell containing a total of 217 atoms. Given the large number of runs needed to perform a statistical analysis, we used the Tersoff-like potential proposed in reference \onlinecite{Pun-2017} for silicon, since it is able to  produce different local minima for the Si interstitial defect. For all cases, as the population founder, we took a configuration that gradient-based optimization methods relax to a minimum of the PES with an energy around 0.5 eV larger than the global minimum. The other parameters were set as follow: $\sigma^{(0)} = 0.08$ \AA, $c_n = 4$  \AA, $c_r = 4$ \AA, $c_\alpha = 0.6$ \AA$^2$/eV.   Different variations of the CMA-ES algorithm are then applied in order to find the global minimum. For each algorithm, 50 independent runs are performed. A run is considered to be converged when the standard deviation of the energy of the individuals within a generation remains below 0.01 eV for at least ten consecutive generations. The converged solution is then relaxed, without any constraint on the atomic positions, applying a gradient-based optimization method. 
 
\begin{figure}
\subfloat{{\includegraphics[width=0.45\textwidth]{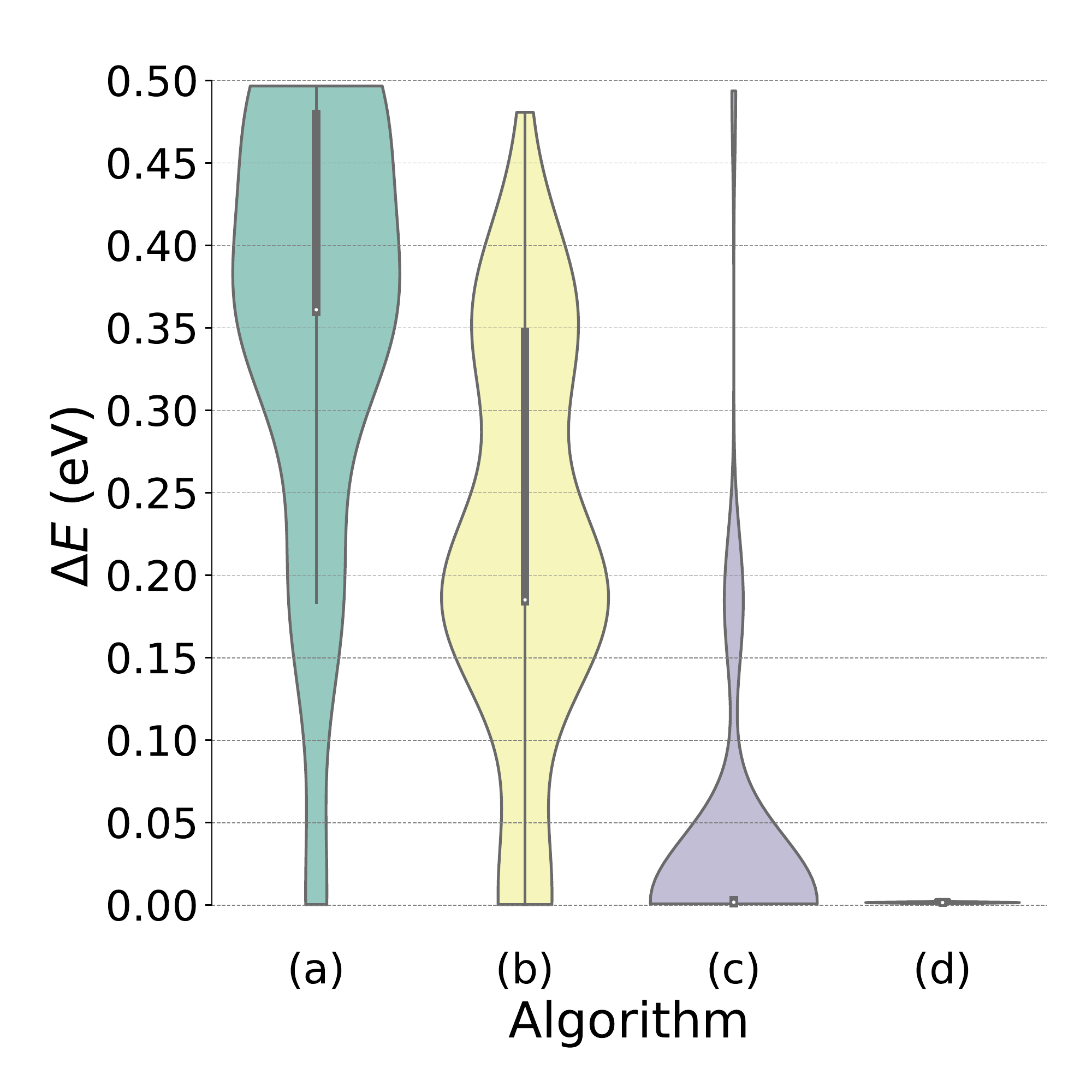}}} 
\subfloat{{\includegraphics[width=0.45\textwidth]{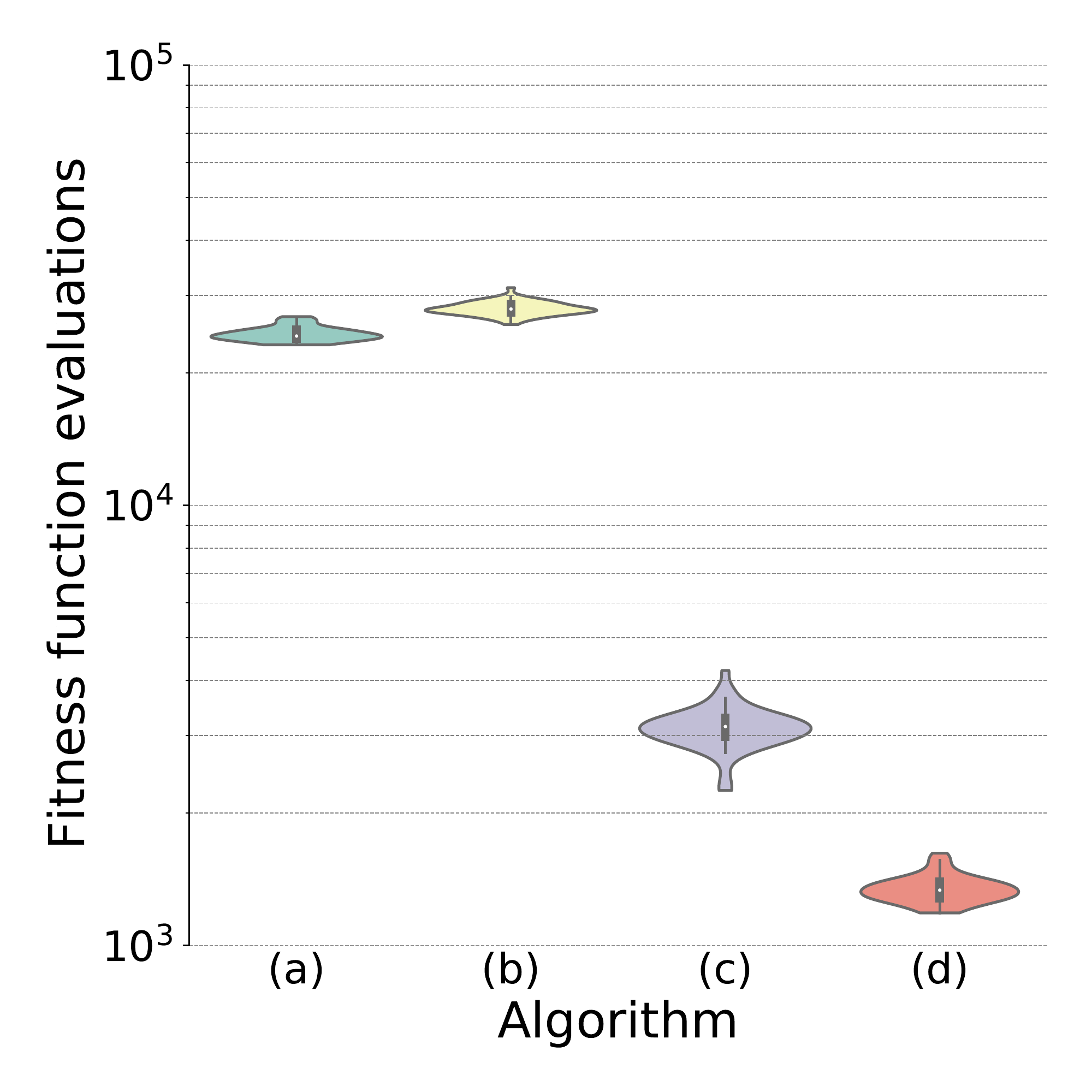}}} %
\caption{\label{fig:algo} Violin plots comparing the performance of different EAs. (a) Original CMA-ES. (b) Initialization of the covariance matrix as per equation \eqref{eq:c_init}, without using a hard cutoff. (c) Initialization of the covariance matrix as per equation \eqref{eq:c_init}, using a hard cutoff. (d) The final version of the algorithm, as proposed in this work including gradients in the update of the distribution mean.
Left panel: energy of the converged solution with respect to the global minimum. Right panel: number of fitness-function evaluations needed to obtain a converged solution.}
\end{figure}
The performance of the algorithms is compared in Figure~\ref{fig:algo}, where the distribution of the energies of the converged solution (with respect to the energy of the global minimum) and the distribution of the number of fitness-function evaluations needed to reach the solution are illustrated. Initializing the covariance matrix according to equation~\eqref{eq:c_init} \emph{without} introducing a hard cutoff (label (b)), already helps the algorithm to converge to the global minimum more often.  A drastic improvement is obtained however when the hard cutoff is also enforced (label (c)), illustrating the detrimental role of a high search-space dimensionality on the overall performances. 
As one can see, the algorithm proposed in this work (label (d)) is both the most robust and requires the least number of fitness-function evaluations, leading to much lower computational costs. As mentioned above, updating the mean using the gradient of the fitness-function might prevent the convergence of the EA, especially when large $\sigma^{(0)}$ and  $c_r $ are used. In this example, only four runs do not converge due to this issue. On the other hand, in all other cases, the EA always converge to some solution.
As shown by this discussion, the modifications applied to the  algorithm lead to fever evaluations of the fitness function and a more robust implementation than the original one (label (a)).

\subsection{Case Study: the Si$_i$ in Si with DFT\label{subsec:Si_int2}}

The short-ranged nature of interactions in the empirical potential used in the previous section produces a too complex PES, with several false minima, as one can notice from Figure \ref{fig:algo}.
We will now illustrate the algorithm performances on the Si$_i$ defect employing DFT. In particular, we performed first-principles calculations using the VASP code \cite{Kresse-1996}, employing the projector augmented-wave (PAW) method \cite{Bloch-1994} and the local-density approximation (LDA) for the exchange-correlation functional \cite{Kohn-1965,Ceperley-1980}. After the EA has converged to a solution, the obtained structure is relaxed to the closest minimum, without any constraint on the atomic positions, using a gradient-based approach until the remaining forces on each atom are less than 0.01 eV/\AA.
We employed a $2\times 2 \times 2$ expansion of the conventional cubic cell, obtaining a supercell containing 64 atoms. The basis set included all plane waves with a kinetic energy within 307 eV and the $3s$ and $3p$ electrons were considered as valence ones. Reciprocal-space integrals were approximated by using the $\Gamma$-point only. 

These computational parameters are adequate enough for demonstrating the capabilities on our proposed EA: they allow for fast calculations and yet are able to reproduce the  defect configurations observed in other DFT studies \cite{Leung-1999,Goedecker-2002,Mattsson-2008,Mariya-2015}. In particular, among the various defect configurations that have been observed in these studies, we consider those three with the lowest energy: the Si in an hexagonal interstitial ($H$), in a tetragonal interstitial ($T$) and the split $\langle 110 \rangle$ configuration, where two silicon atoms form a dumbbell configurations along the  $\langle 110 \rangle$ direction ($X$). Well-converged simulations tend to agree that the $H$ and $X$ configurations have very similar energies, with the $X$ one most likely being the ground state. On the other hand, the energy of the $T$ configuration is noticeably larger \cite{Leung-1999,Mattsson-2008,Mariya-2015}. In the 64-atoms supercell employed in this work, we also found that the $H$ and $X$ configurations have  very similar energies. The ground state is represented by $H$, with the energy of the $X$ configuration being just 23 meV higher. The energy of the $T$ configuration is around 0.37 eV above the ground state.  These results are adequate for the purpose of presenting the method proposed in this work, as all the three low-energy minima are present and the multimodal PES offers a realistic test case for our approach.

We choose a $T$ insterstitial as the population founder. While a gradient-based optimization would lead to the $T$ configuration, our EA correctly converges to the $H$ one. In particular, we repeated the simulation three times, each time setting $c_n = 4 $ \AA. In two calculations $c_\alpha^{(0)}$ was set to 0~\AA$^2$/eV and in one to 0.3 \AA$^2$/eV. For the former set of calculations, $\sigma^{(0)}$ was set to  0.2~{\AA}, for the latter to 0.08 {\AA}. Finally, $c_r$ was set either 0, or 10 {\AA}. In all  cases, the algorithm converged to the ground state.

As mentioned in section~\ref{subsec:EA}, a balance between  $\sigma^{(0)}$, $c_\alpha^{(0)}$ and $c_r$ must be kept. As too large initial variances will lead to highly disordered structures characterized by large energies. Hence, if $c_\alpha^{(0)} \neq 0$, the population mean will be shifted to areas of the PES far from any minimum, making the convergence very slow and diminishing the probability to find the global minimum. In general, one should reduce $c_\alpha^{(0)}$ the larger $\sigma^{(0)}$ and $c_r$ are.
However, we can use this phenomenon in order to force the algorithm to converge to some local minimum. To give an example, the algorithm converges to the $X$ configuration by setting  $\sigma^{(0)}$ to 0.15 {\AA}, $c_\alpha^{(0)}$ to 0.05 \AA$^2$/eV and $c_r$ to 25 {\AA}. Additionally, another local minimum with energy of around 0.4 eV above the ground state was found by increasing $\sigma^{(0)}$ to 0.18 {\AA}.
This discussion shows how an extensive exploration of the PES can be obtained by increasing systematically the algorithm parameters controlling the variance for atomic displacements, forcing the algorithm to converge not only to the global minimum but to local ones as well.
As the evolutionary process is stochastic in nature, results will generally vary from run to run. However, the above analysis describes a trend we observed all the times our method was employed.

\begin{figure}
\centering
\includegraphics[width=0.9\textwidth]{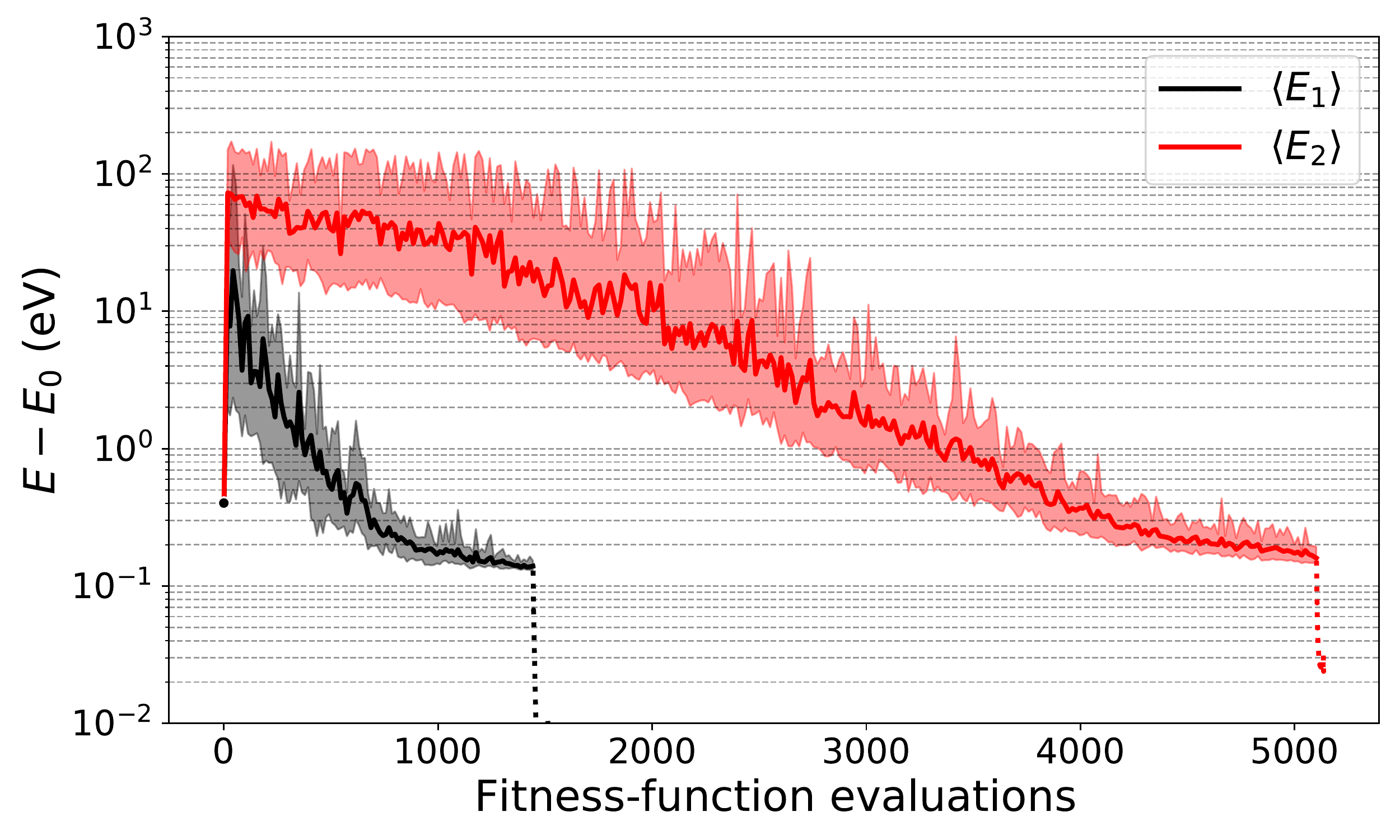}
\caption{\label{fig:runs_Si} Average population energy as a function of the number of fitness-function evaluations for the Si$_i$ in the LDA calculations. The zero of the energy is set to the ground state configuration for this defect ($H$). The maximum and minimum energies for each generation are also shown. Black: initial parameters: $c_n = 4 $ \AA,  $c_r = 10$ {\AA},  $\sigma^{(0)} = 0.08$ {\AA} and $c_\alpha^{(0)} = 0.3$ \AA$^2$/eV. Red:  initial parameters: $c_n = 5 $ \AA,  $c_r = 25$ {\AA},  $\sigma^{(0)} = 0.15$ {\AA} and $c_\alpha^{(0)} = 0.05$ \AA$^2$/eV. The dotted lines represent the calculations needed to relax the converged solution to the equilibrium configuration using gradient-based optimization.}
\end{figure}
Figure~\ref{fig:runs_Si} shows the average population energy as a function of the number of evaluations of the fitness function. The zero of the scale is set to the energy of the fully relaxed $H$ configuration, which represents the ground state. The bold black line represents a run that converges to this ground state. Note that since a hard cutoff is applied, the energies of the converged structures will be larger than the ground state one. In any case, if the cutoff is large enough, a gradient-based optimization can be used to quickly relax these structures to the proper global minimum, as shown by the dotted lines. One can also note that in the last generations, the average population energy changes very little. In the present work, we consider the algorithm to be converged when the population standard deviation remains below 0.01 eV for at least ten consecutive generations. This in an extremely conservative criterion: for this system, setting a threshold of 0.1 eV for the population standard deviation still yields a converged solution that relaxes to the ground state, but requires about half of the number of fitness-function evaluations.  Another run of the EA is shown in Figure \ref{fig:runs_Si} by the red line. In this run we used the initial parameters that lead the EA to converge to the $X$ configuration. Both runs start from the same population founder, but, by comparing the two lines, is clear that larger values of $c_\alpha^{(0)}$ and $\sigma^{(0)}$ yield structure with a much higher energy, making the algorithm convergence slower and reducing the probability of finding the global optimum.

\subsection{Unsupervised Machine Learning Model \label{subsec:CLUS}}
While the CMA-ES algorithm has proven to be very reliable on a large set of non-separable multimodal benchmark functions, in some complex cases, particularly large population sizes are required in order to reach the global optimum  \cite{Hansen-2004}. Moreover, even when the algorithm converges to the global optimum, it might also sample basins containing other low-energy minima, a phenomenon which is of interest for finding  metastable defect configurations. As discussed in Section~\ref{subsec:Si_int2}, it is possible to find metastable structures by varying the initial parameters, but it is necessary to rerun the algorithm in order to make it converge to these competing minima. While this offers a simple and systematic method for finding metastable structures, one would wish to limit the number of DFT calculations.   A method to exploit all the data generated during an evolutionary run and discover low-energy minima without the need to re-run the algorithm would thus be particularly useful.

\begin{figure}
\centering
{{\includegraphics[width=.6\textwidth]{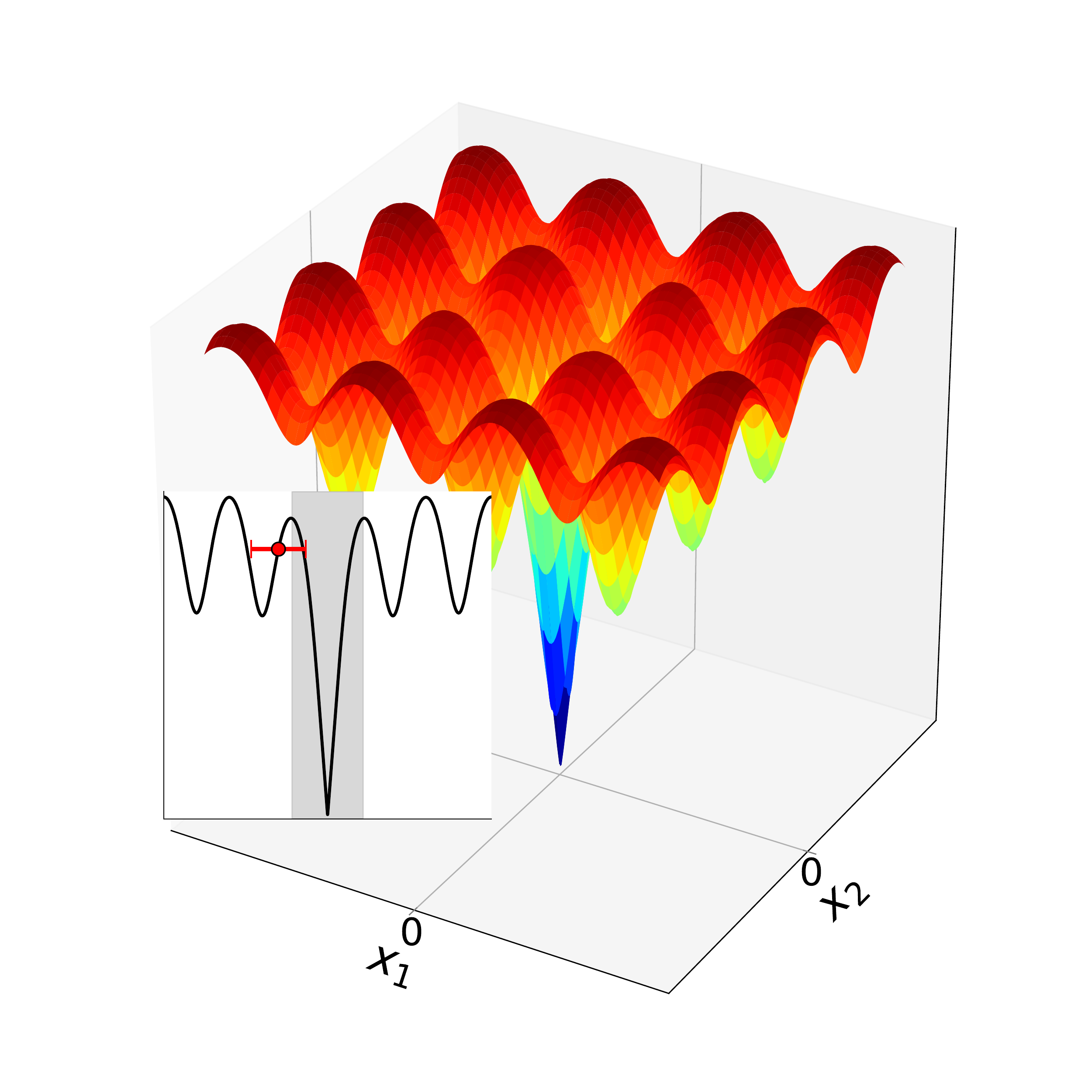}}}
\caption{\label{fig:ackley} A reparametrization of the two-dimensional Ackley function, $f(x_1,x_2)=-a\exp \left(-b{\sqrt {\frac{1}{2}\left(x_1^{2}+x_2^{2}\right)}}\right)-\exp \left[\frac{1}{2}\left(\cos(c x_1)+\cos(c x_2)\right)\right]+e+a$, with $a = 3$, $b = 1$, $c = \pi/3$, chosen in such a way that the basins of the local minima and of the global minimum have a comparable width. In the inset, the function is evaluated on the $x_2 = 0$ plane. The red dot represents the location of $\mathbf{m}^{(0)}$ and the red error bar represents the magnitude of $\sigma^{(0)}$. The gray area represents a region of the search space where a gradient-based optimization approach would lead to the global minimum.}
\end{figure}
To illustrate the ability of the EA to explore basins of the PES different from the one containing the converged solution, we take a multimodal function whose global minimum is difficult to reach,  Figure~\ref{fig:ackley}, as an example. Even though the cost function at the global minimum, $\mathbf{x}^\star$, has a considerable lower value than the local minima, convergence to $\mathbf{x}^\star$ is not straightforward. As a concrete example, consider the situation depicted in the inset. The red dot represents the location of the CMA-ES initial mean $\mathbf{m}^{(0)}$ and the red error bar represents the magnitude of $\sigma^{(0)}$. We employ the basic implementation of the CMA-ES and use a population size of 20 individuals. In such conditions, even though the initial global step-size is large enough to allow exploration of the basin of $\mathbf{x}^\star$, the algorithm might not converge to this optimum, as the region of the $\mathbf{x}^\star$ basin where cost function is lower than other minima is relatively narrow. As a consequence, the update of $\mathbf{m}$ might lead the individuals distribution to shift toward other basins. As a matter of fact, if we define $p$ as the probability that the CMA-ES will converge to $\mathbf{x}^\star$ in this problem, and let $\hat{p}$ be the maximum likelihood estimator of $p$, then, after running the algorithm 3000 times, the calculated value of $\hat{p}$ is found to be $\approx 0.79$. This is a rather low probability, considering the fact that the problem is only two dimensional. For comparison, if one employs the standard parametrization of the Ackley function, keeping the other parameters unaltered, the calculated $\hat{p}$ turns out to be $1$.
Nevertheless, even when the algorithm does not converge to  $\mathbf{x}^\star$, it still visits regions of the search space within its basins. For this problem, we found indeed that the estimated expected number of times this happens is around $10$. Even though only a very small fraction of individuals visits the basin of the global minimum, in the context of the present work  we might expect that structures located in the global minimum basin to be noticeably different from those located in other basins, and thus recognizable by some unsupervised learning method even if there are very few of them.

In particular, analyzing the way the CMA-ES operates in the evolutionary process, one would expect that in the first generations widely different structures are generated. As the evolution proceeds, the algorithm will tend to converge to some basin, and the overall covariance of the individuals distribution will decrease, generating more and more similar structures. Assuming that the structures belonging to a certain basin are characterized by certain common features, employing appropriate structural descriptors these structures will appear as localized clusters. This suggests using a density-based clustering algorithm in order to identify these structures, as described in the example below.

\subsection{Case Study: Pristine silicon with DFT \label{subsec:ffcd}}

For presenting the unsupervised model used to find low-energy defect configurations, pristine silicon is an interesting case study. For this system, the global minimum is obviously the defect-free diamond lattice structure; however there are several meta-stable defect laden structures, including Frenkel pairs, where a Si moves to an interstitial position and leaves behind a vacancy, and the four-fold coordinated defect (FFCD)\cite{Cargnoni_PRB98,Goedecker-2002}, which is possibly the intrinsic defect in silicon with the lowest formation energy. 
We considered a pristine silicon supercell containing 64-atoms, centering the "defect position" to one of the Si atoms and using the computational parameters: $c_n = 4 $ \AA,  $c_r = 20$ {\AA},  $\sigma^{(0)} = 0.1$ {\AA} and $c_\alpha^{(0)} = 0.05$ \AA$^2$/eV. Running the EA, we found it to converge, as expected, to the diamond structure.


\begin{figure}[t]
\centering
\includegraphics[width=.9\textwidth]{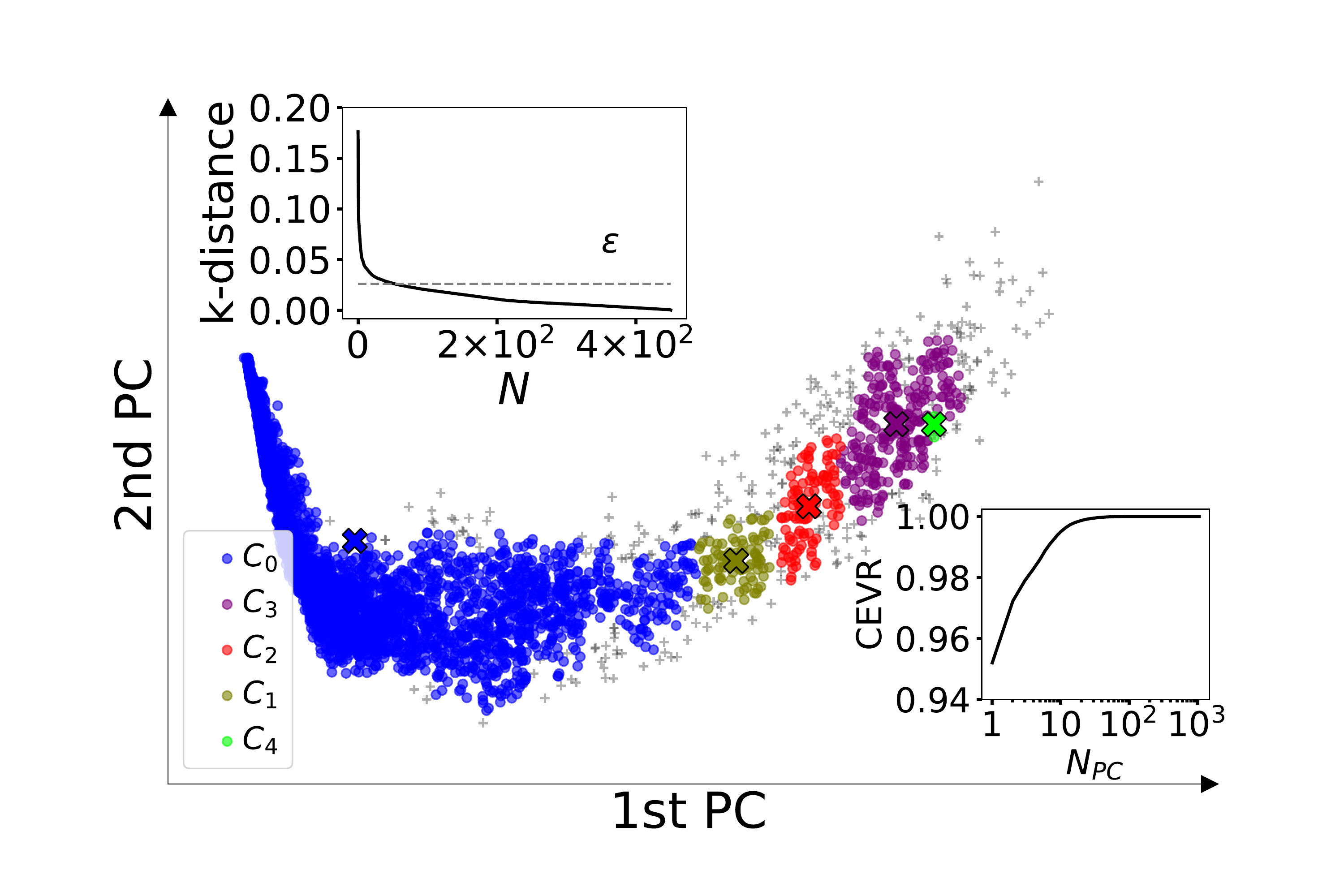}
\caption{\label{fig:si_clusters} Representation of the structures visited by the evolutionary algorithm for pristine bulk silicon crystal. Scatter plot of the first two principal components of crystal fingerprint descriptors for the structures generated during the EA run  (see main text for the details).  Cluster centroids are shown with crosses. The inset at the top-left corner shows the $k$-distance plot used to find the clustering parameter $\varepsilon$. The inset at the bottom-right corner shows the cumulative explained variance ratio (CEVR) as a function of the number of principal components ($N_{PC}$). Cluster names are given according to their distance from the left-most cluster and they are ordered in the legend according to their size.  }
\end{figure}
In order to exploit all the data generated during this evolutionary run, we represented all the generated structures using the crystal structural fingerprints proposed by Oganov and Valle \cite{Valle-2010}. We then performed principal-component analysis (PCA) and kept only the first two components, which preserve more than 95 \% of the dataset variance,  as shown in the bottom-right inset of Figure~\ref{fig:si_clusters}. The main panel of Figure~\ref{fig:si_clusters} shows the first two principal components of the collected structure. The shape of the scatter plot is very characteristic and we can compare it to that of a comet: a dense region of structures, those obtained in the last generations, form the head, while the tail is more rarefied and is composed by the structures generated in the first generations. The first principal component strongly anticorrelates with the generation number so that the further a structure appears to the left of the plot, the more it belongs to later generations.
As expected, between the head and the tail one can notice some regions with a noticeably higher density of structures. In order to collect them into distinct clusters, we employed the density-based \texttt{DBSCAN} algorithm \cite{Ester-1996}. This algorithm uses core points in order to identify clusters. A core point $q_C$ is a point such that a sphere of radius $\varepsilon$ centered on it contains at least $minPts$ points. All points with a distance of less than $\varepsilon$ from $q_C$ are said to be directly density-reachable from it. A cluster is made of all points for which a chain of sequentially directly density-reachable points connecting to a core point exists. All other points are considered to be noise and represented by crosses in Figure \ref{fig:si_clusters}. To obtain this figure, we set $minPts$ to 8 and the value of $\varepsilon$ was chosen by analyzing the 8-distance plot shown in the top-left inset of the figure. In particular, the exact value of  $\varepsilon$ has been chosen in order to obtain distinct and dense clusters.

\begin{figure}
\centering
\includegraphics[width=0.5\textwidth]{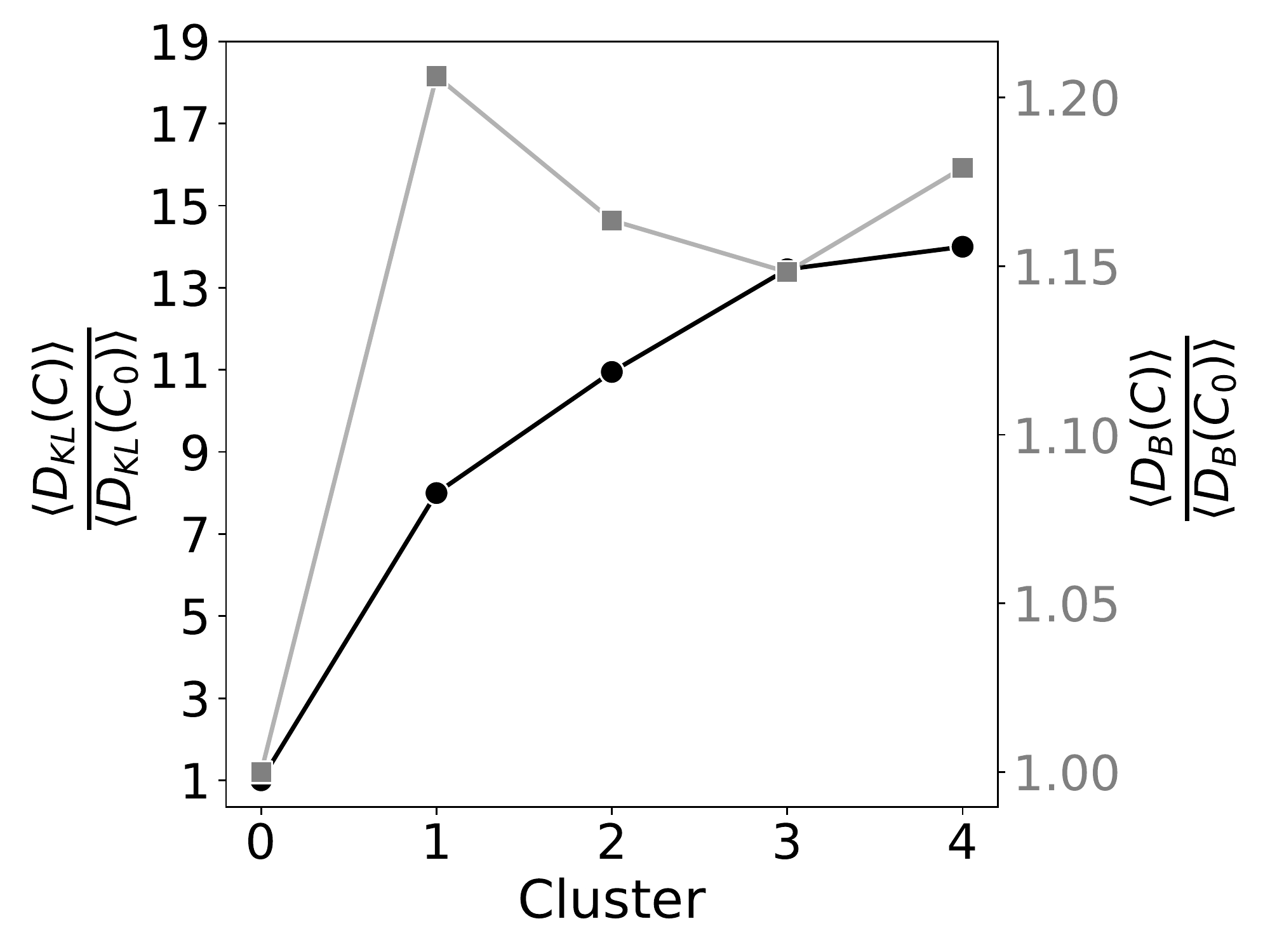}
\includegraphics[width=0.3\textwidth]{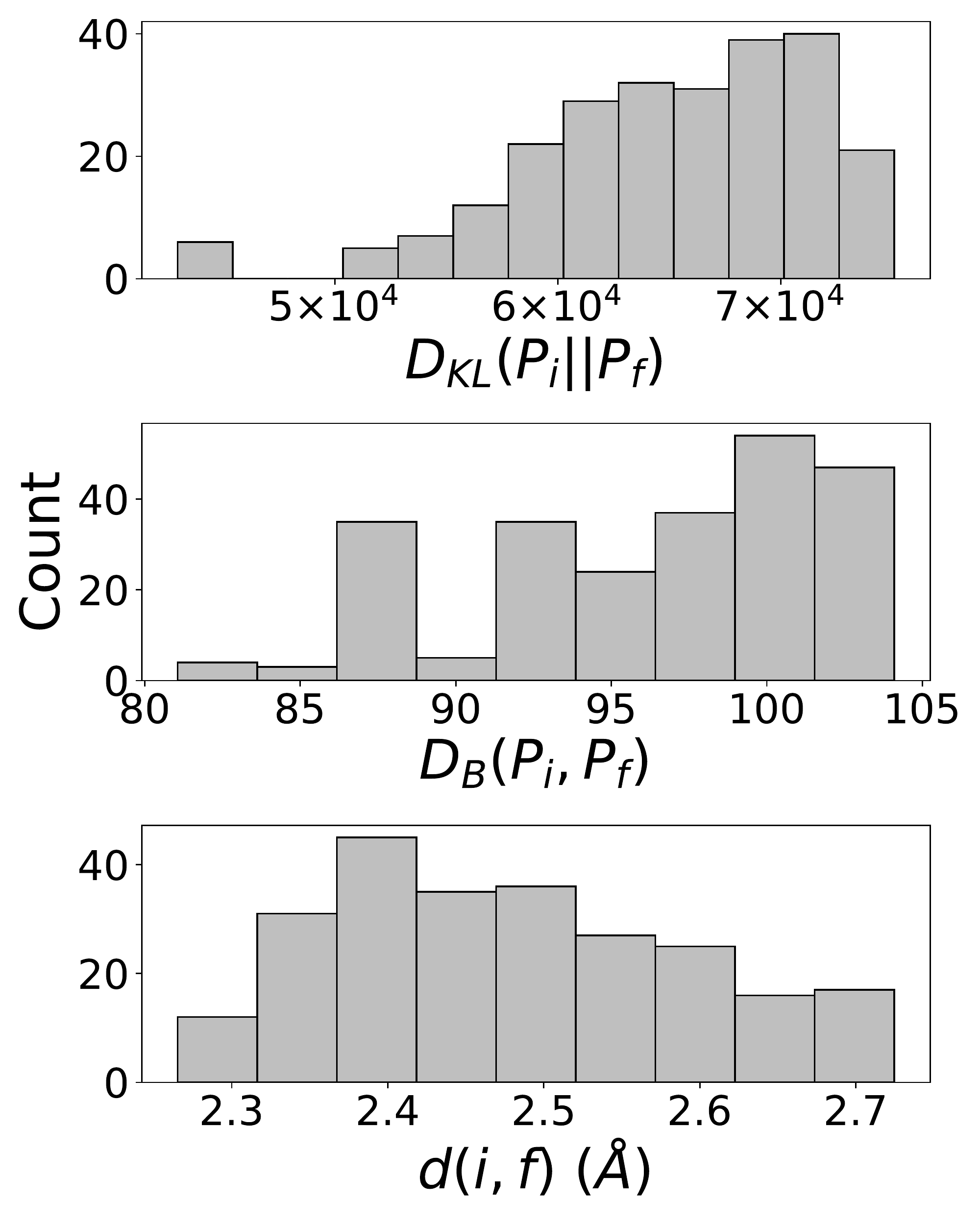}
\caption{\label{fig:si_clusters2}  Left: cluster relative average KL divergence and relative average Bhattacharyya distance. Right: distribution of the KL divergence,   Bhattacharyya distance and Euclidean distance for the individuals in cluster $C_3$. }
\end{figure}
The obtained clusters were labeled according to their Euclidean distance from the converged solution (the pristine Si supercell), with $C_0$ being the cluster containing the converged structure and $C_4$ being the cluster whose centroid is the most distant from that of $C_0$.
 The Euclidean distance perhaps also offer he most natural way of selecting the most interesting clusters. This offers a measure of how distinct structures differ according to their geometric representation.  Additionally, one can also consider distances in distributions, specifically how much the distributions that have generated the individuals in the clusters of interest differ on average from the distribution of the converged solution. In order to assess the similarity between two probabilities distributions, we use two  quantities: the Kullback-Leibler (KL) divergence and the Bhattacharyya distance. The KL divergence quantifies the similarity between probability distributions $P$ and $Q$ according to:
\begin{equation}
\label{eq:KL}
D_{KL} \left( P || Q \right) = \text{E}_{X \sim P} \left[ \ln \frac{P(X)}{Q(X)}  \right],
\end{equation}
where $X$ indicates a random variable and the expectation is taken with respect to $X$ which  is distributed according to $P$.
The Bhattacharyya distance function takes the form:
\begin{equation}
\label{eq:batta}
D_B(P, Q) = -\ln \bigg( \int\!\! \sqrt{P(x) Q(x)} dx \bigg)
\end{equation}

Figure~\ref{fig:si_clusters2} shows the clusters relative average KL divergences and Bhattacharyya distances. In particular, the cluster average divergence or distance is calculated according to:
\begin{equation}
\langle D(C) \rangle = \frac{1}{n_C} \sum_i^{n_C} D(P_i, P_f)
\end{equation}
where $P_i$ represents the distribution of the population where structure $i$ belongs, and $P_f$ represents the distribution of the last population and $n_C$ the number of structures in that particular cluster.

One can see that in this case the relative KL divergence strongly correlates with the distance from $C_0$. On the other hand, the average  Bhattacharyya distance noticeably differs from the value obtained for $C_0$ but does not differ too much among the other clusters. In any case, cluster $C_4$ seems to be an ideal candidate for discovering new structures as it is the most different in geometric representation, as quantified by the centroid Euclidean distance, and also in distribution, according to the KL divergence, from the converged solution. $C_4$ contains only three structures, optimizing them to the closest minimum using a gradient-based method results in a structure relaxing back to the pristine system, one relaxing to the FFCD and one to another defective configuration with a much larger formation energy: around 6.7 eV. 

\begin{figure}
\centering
\includegraphics[width=1\textwidth]{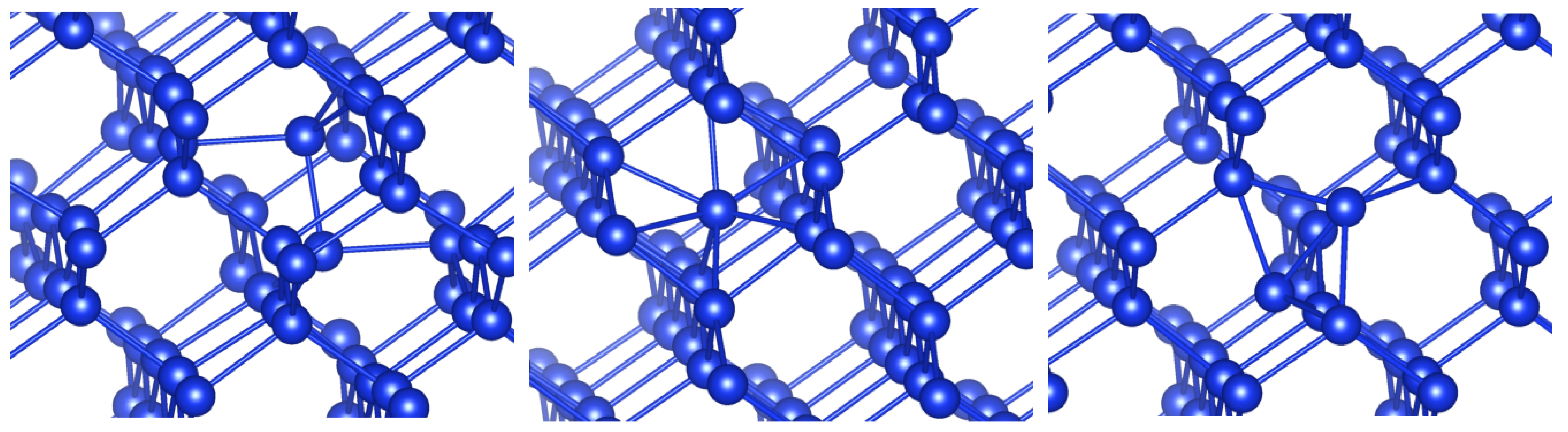}
\caption{\label{fig:geometries_si} Defective structures in stoichiometric Si supercells found by the clustering method. Left: the FFCD. Middle: the Frenkel pair where  Si$_i$ is in an hexagonal interstice. Right: the Frenkel pair where the interstitial Si is in the split $\langle 110 \rangle$ configuration. Images created with \texttt{VESTA} \cite{Momma-2011}.}
\end{figure}
Since $C_4$ contains very few structures, one would want to consider $C_3$ as the next cluster of interest. $C_3$ contains almost 250 structures and it is not convenient to optimize all of them. One has to employ a finer selection scheme for shortlisting them.  Following the idea of choosing structures according to their Euclidean distance and distribution distance with respect to the converged solution, in Figure~\ref{fig:si_clusters2} we show the distributions of these quantities for the structures in $C_3$. As the clustering algorithm is based on the density in the linear space spanned by the first two principal axes, it is not surprising that the variance of the Euclidean distance from the converged solution is relatively small. However, structures sampled from the same distribution can be rather different; therefore the same cluster might show a much larger variance for the KL divergence and Bhattacharyya distance. Hence, we selected only those structures in $C_3$  whose Euclidean distance from the converged solution falls in the upper quartile range and whose KL divergence and Bhattacharyya distance are above the respective median values. This procedure returns 30 structures, which are then optimized with a gradient-based method. Of these, 22 relax back to the pristine supercell, one relaxes to the FFCD, two structures relax to a defective configuration with a formation energy of around 4.4 eV, which can be described as a Frenkel pair where the interstitial atom is in a $H$ site, one structure relaxes to another Frenkel pair, this time the interstitial is in a $X$ site and has formation energy of around 4.6 eV, and the remaining 4 structures relax to more disordered configurations with formation energies between 6.4 and 9.5 eV. The three discovered defective structures with the lowest formation energy are shown in Figure \ref{fig:geometries_si}. While different filtering criteria than the one suggested here can be used,  this analysis shows the efficacy of the presented approach for selecting a small number of structures which lead to diverse low-energy minima with a modest computational cost.

\subsection{Supervised Machine Learning Model \label{subsec:GPR}}
The algorithm outlined in section \ref{subsec:EA} is in itself able to find low-energy defect configurations, for all systems considered in this work, at a reasonable computational cost. Nevertheless, the convergence of the self-consistent field cycle might pose a more daunting task in terms of time and computer power for particular  systems and/or when very large supercells must be employed. 
It is therefore desirable to use a ML metamodel able to accurately predict  the system electronic energy and thus reduce the number of required first-principles calculations and hence the computational burden.
As we mentioned in the introductory section, efficient materials descriptors and models have been shown to be able to approximate DFT energies within few meV per atom  \cite{Behler-2007,Bartok-2010,Rupp-2012,DeVita-2015,Bartok-2018}. Among the various proposed models, we have chosen to employ GP regressors \cite{Rasmussen-2006}, as they have already shown to be very effective in reducing the amount of first-principles calculations required in evolutionary algorithms \cite{Jennings-2019,Bisbo-2020}. Furthermore, the underlying Bayesian formalism enables the use of an  on-the-fly active learning training during the evolutionary process, as will be described later in this section.

A GP is a collection of random variables such that any finite subset thereof is a jointly-Gaussian random vector. In our case, we are interested in modelling the cost function $f$ as a Gaussian random variable. The joint distribution of these random variables in uniquely defined by a mean function $m(\mathbf{x})$ and covariance, or kernel, function $k(\mathbf{x}, \mathbf{x'})$. In particular, let $\mathbf{X}^*$ be a $n_*\times d$ matrix where each row represents an input vector at which one would want to predict the value of $f$, then the joint prior distribution of the random vector $\mathbf{f}^*$, where $f^*_i = f(\mathbf{x}_i^*), \, \, i= 1, \dots, n_*$, is, according to the model:
\begin{equation}
\label{eq:prior}
\mathbf{f}^* | \mathbf{X}^*, m, k \sim \text{N} \left(\mathbf{m}^*, K \left(\mathbf{X}^*, \mathbf{X}^*\right) \right),
\end{equation} 
where the components of $\mathbf{m}^*$ are given by $m_i^* = m(\mathbf{x}_i^*)$ and the elements of the covariance matrix $K$ are given by: $K_{ij} = k(\mathbf{x}_i^*, \mathbf{x}_j^*)$.
If one has observed values of $f$ on a set of data points $\mathbf{X} \in {\rm I\!R}^{n \times d}$, then this information can be used to obtain the joint posterior distribution of $\mathbf{f}^*$:
\begin{equation}
\label{eq:posterior}
\mathbf{f}^* | \mathbf{X}^*, \mathbf{X}, \mathbf{f}, m, k \sim \text{N} \left(\bm{\mu}^*, \bm{\Sigma}^*\right),
\end{equation} 
where the posterior mean $\bm{\mu}^*$ and covariance matrix $\bm{\Sigma}^*$ are given, respectively, by:
\begin{gather}
\label{eq:posterior_mean}
\bm{\mu}^* = \mathbf{m}^* + K\left(\mathbf{X}^*, \mathbf{X}\right) K^{-1}\left(\mathbf{X}, \mathbf{X}\right) (\mathbf{f} - \mathbf{m}) \\
\label{eq:posterior_cov}
\bm{\Sigma}^* = K \left(\mathbf{X}^*, \mathbf{X}^*\right) - K\left(\mathbf{X}^*, \mathbf{X}\right) K^{-1}\left(\mathbf{X}, \mathbf{X}\right) K\left(\mathbf{X}, \mathbf{X}^*\right).
\end{gather}
Once the posterior of $\mathbf{f}^*$ is obtained, one can make predictions at inputs $\mathbf{X}^*$ by using the posterior mean $\bm{\mu}^*$: the estimator which minimizes the mean squared error of the prediction. The most critical aspect affecting the posterior distribution is the choice of the covariance function $k$. Several kernels have been proposed in the literature. In the present work, we employ a linear combination of two squared exponential kernels:
\begin{equation}
\label{eq:kernel}
k(\mathbf{x}, \mathbf{x'}) = C \left(x \exp{-\frac{||\mathbf{x} - \mathbf{x'}||^2_2}{2l_1^2}} + (1 - x) \exp{-\frac{||\mathbf{x} - \mathbf{x'}||^2_2}{2l_2^2}} \right).
\end{equation}
The length scale hyperparameters $l_1$ and $l_2$ have been constrained on the range $\left[10^2, 10^6 \right]$ and $\left[10^{-2}, 10^2 \right]$, respectively, in order to add flexibility to the prediction distribution. A similar kernel has been used in reference \onlinecite{Bisbo-2020} in an GA context and has shown good performances. The values of the hyperparameters $C, l_1$ and $l_2$ are obtained during the training phase by maximizing the log marginal likelihood, as described in reference \onlinecite{Rasmussen-2006}; while the value of $x$ is chosen by the user in $[0, 1]$. Regarding the mean function, we took it as the sample mean of the dataset. 

\begin{figure}[t]
\includegraphics[width=.9\textwidth]{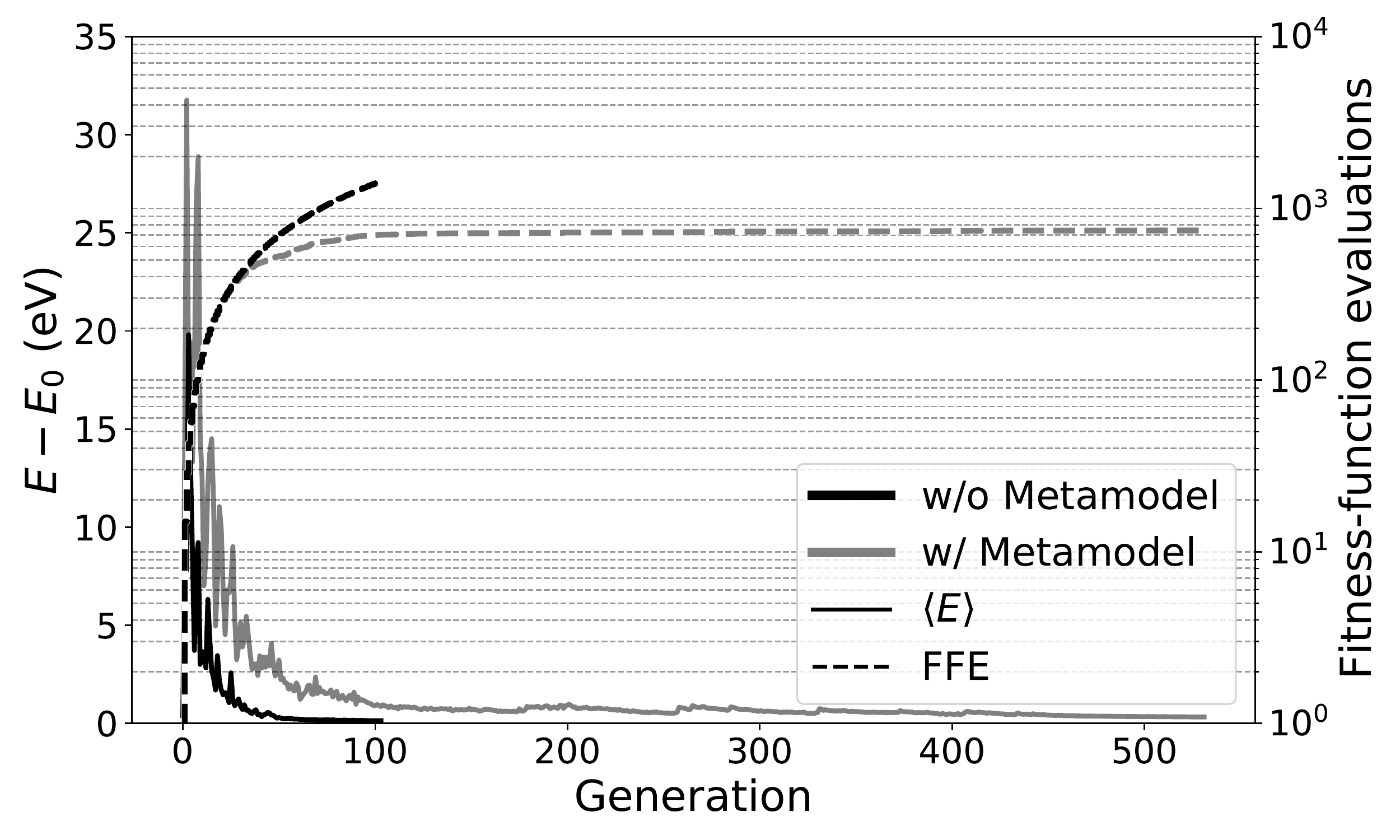}
\caption{\label{fig:meta} Comparison between two runs of the EA for the Si$_i$ defect. One is performed using only DFT data (black lines) and one is performed with the use of the metamodel (gray lines). Dashed lines represents the number of required fitness-function evaluations. The bold lines, the population average energy referenced to the ground state energy of Si$_i$. LDA calculations.}
\end{figure}

An advantage of having the posterior covariance matrix, is that for each prediction we have an estimation of its uncertainty. We use this quantity in order to decide which structures generated by the EA should have their fitness evaluated directly by a DFT calculation, or approximated by the GP metamodel. In particular, at each generation $g$, the GP metamodel is trained using the individuals in the previous generations whose energy has been calculated directly by DFT. The model then calculates the prediction standard error for each individual in generation $g$. If this quantity is larger than a user-define threshold, then the DFT energy is calculated for that individual, otherwise the GP-predicted energy is used. This is an effective way to train the metamodel on-the-fly allowing for a certain degree of active learning: the metamodel can automatically choose the proper training data needed to maintain a predefined accuracy for the approximated PES. It is important to note that it is not necessary for the latter to be an exact reproduction of  the true PES: the approximation should be good enough to represents the main basins of the PES, but these need not to be an exact replication of the real ones. The main purpose of the metamodel is to offer a surrogate PES which can be used for sampling the real one at a much lower computational cost. 
With the metamodel in use, the number of fitness-function evaluations can be considerably reduced, as shown in Figure~\ref{fig:meta}. In particular, we considered the same Si$_i$ defect and computational parameters as in Section~\ref{subsec:Si_int2}. Two runs were performed: one run employed the metamodel during the evolutionary process while the other run did not. In both cases, the converged solution relaxed to the ground state. The metamodel was trained employing the crystal fingerprint descriptors proposed by Valle and Oganov \cite{Valle-2010}. The minimum standard error allowed for the metamodel was set to 10 meV/atom. From Figure~\ref{fig:meta}, one can notice that when the metamodel is employed, no explicit energy calculations need to be performed after a certain amount of generations. Overall, even tough the metamodel run needs a larger number of generations in order to converge (partly because no gradient term is used in updating the population mean), it requires a considerable less number of  fitness-function evaluations. In particular, this number is reduced by more than a factor of two in this example.

\section{\label{sec:Results} The neutral $\square_\text{O}$ in TiO$_2$ anatase}
\begin{figure}[t]
\centering
\subfloat{{\includegraphics[width=0.25\textwidth]{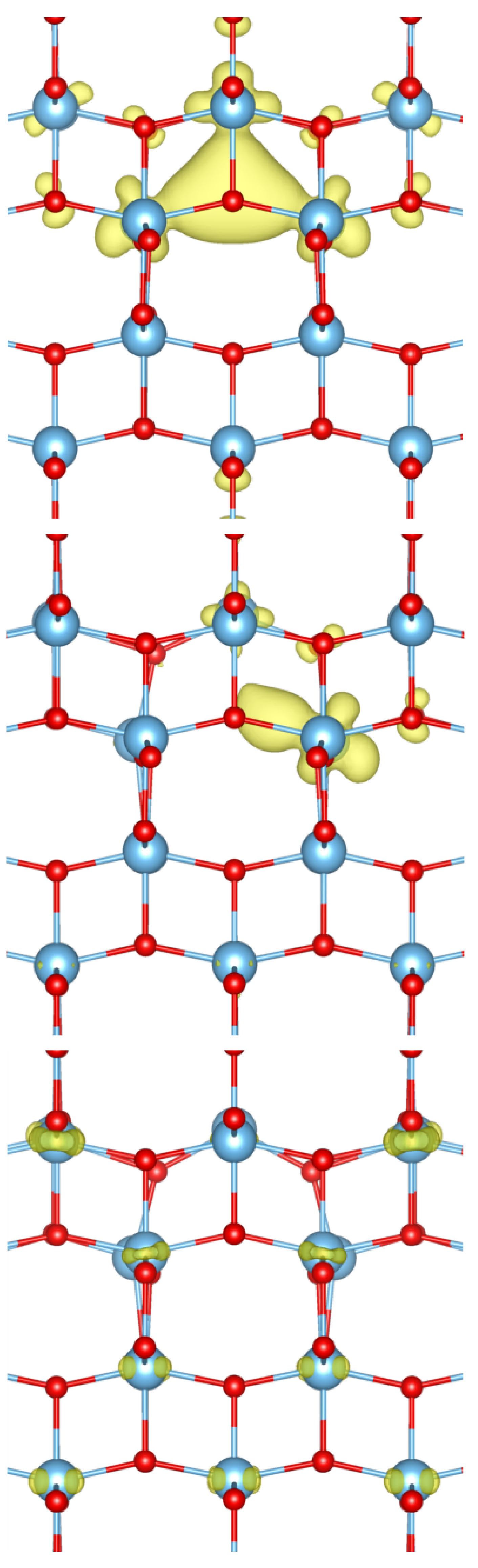}}}
\subfloat{{\includegraphics[width=0.4\textwidth]{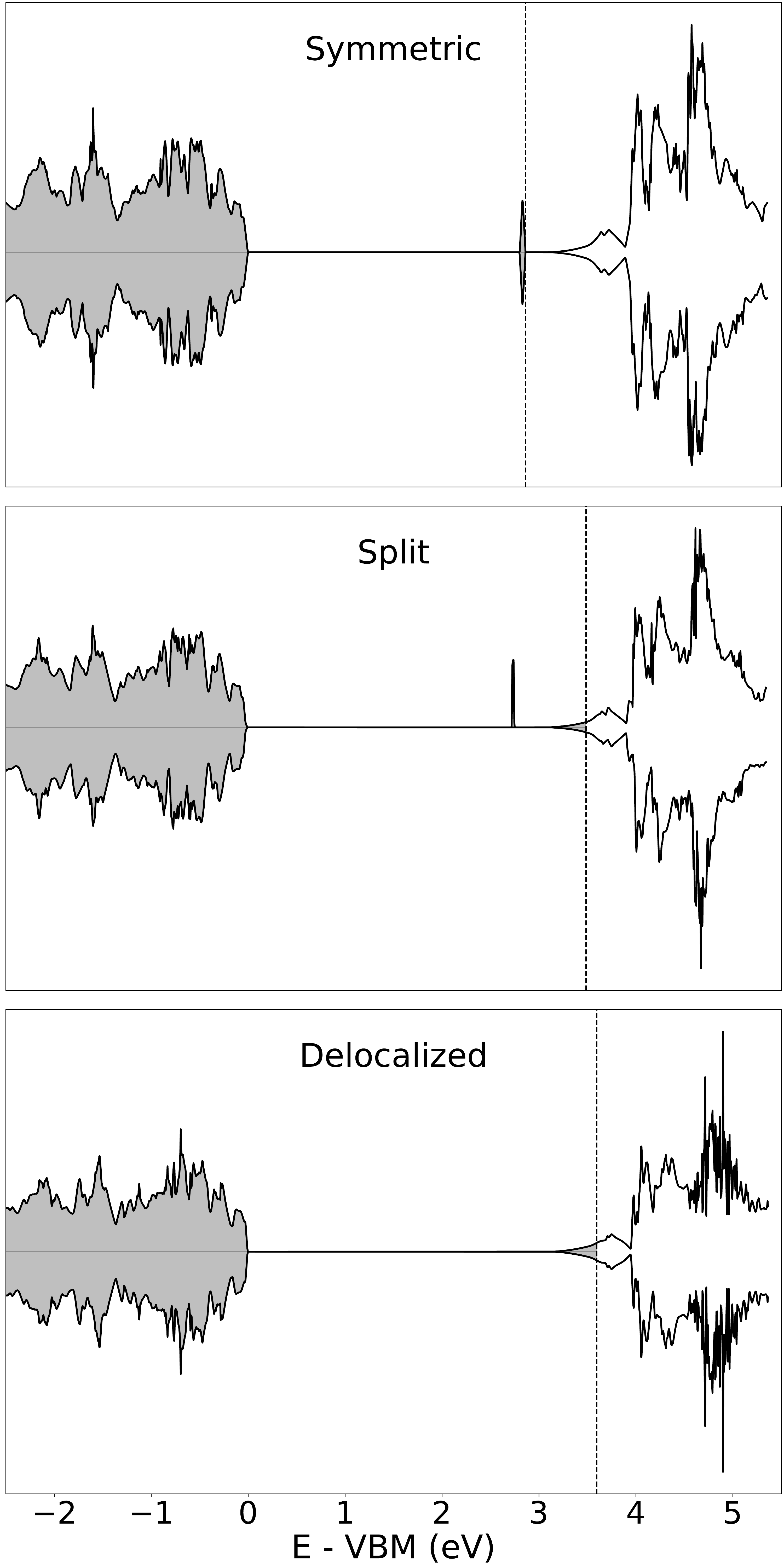}}} %
\caption{\label{fig:geometries} Left panel: Configurations for the $\square_\text{O}$ defect in anatase TiO$_2$ representing PES minima at the HSE15 level. The isosurfaces represent the valence electron density projected onto the defect-induced level. Isosurfaces levels are shown at 0.005 e/\AA$^3$. From top to bottom: the simple vacancy, the split vacancy and the delocalized vacancy. Right panel: electronic density of states for these configurations, calculated at the HSE15 level using a 2$\times$2$\times$2 k-point mesh. The dotted lines represent the Fermi level. The gray areas emphasize occupied states. The energy is referenced with respect to the valence band maximum (VBM).}
\end{figure}

We finally applied the method proposed in this study to a complex defect in a transition-metal oxide.
The uncharged oxygen vacancy, $\square_\text{O}$, is perhaps the native defect of anatase TiO$_2$ that has received the most attention both from experimental and computational studies. Several first-principles calculations have been performed using a wide range of functionals and different structures have been reported. Simply removing an oxygen atom from the host material and relaxing the atomic positions typically leads to the "simple" vacancy structure, which is characterized by the localization of the two extra electrons on the vacancy site and has the point group symmetry $mm2$, top plot in Figure~\ref{fig:geometries}. However, another configuration with considerable lower energy can be obtained. This is called the "split" vacancy configuration and is distinguished by a broken symmetry (point group $\bar{2}$) and the localization of an extra electron on a neighboring Ti atom \cite{Finazzi-2008,Mattioli-2008,Morgan-2010,Arrigoni-2020}, middle of  Figure~\ref{fig:geometries}. Due to the inability of standard DFT to treat strongly localized electrons, the split configuration is most prominent using either the DFT+$U$ approach or hybrid functionals. We have recently shown that DFT+$U$, which is computationally much less demanding, is able to yield geometric structures in good agreement with those obtained with hybrid functionals \cite{Arrigoni-2020}. The agreement of the defect formation energy is considerably worse: using DFT+$U$, we found that the simple vacancy configuration has energy 0.77 eV larger than the split vacancy, while using HSE, this difference is of about 0.38 eV \cite{Arrigoni-2020}. In both cases these minima are however clearly distinguishable.
 Based on these observations, we will explore the low-energy configurations of this defect at the DFT+$U$ level, using the EA and the density based clustering  method in order to explore the PES. A final set of selected structures is then studied also with a hybrid functional.

\subsection{\label{sec:Calculations} Computational Details }
The vast majority of calculations performed for TiO$_2$ in this work employ the DFT+$U$ method \cite{ldaUfll}, where a $U$ of 5.8 eV is applied to the $d$ electrons of the Ti atoms, keeping the same computational parameters described in reference  \onlinecite{Arrigoni-2020}, with the exception that during the EA run, only the $\Gamma$-point is used. After the EA has converged a $2 \times 2 \times 2$ k-point mesh is used in order to relax the obtained solution to the closest minimum. Finally, for selected structures, we performed also calculations employing the Heyd-Scuseria-Ernzerhof screened hybrid functional (HSE) \cite{Heyd-2003}. The amount of Hartree-Fock exchange in the HSE functional was set to 15 \% as described in reference \onlinecite{Arrigoni-2020} (HSE15). In all cases a supercell with 108 atoms was used.
The results we present below are obtained by setting the following values for the EA hyperparameters:  $c_n = 6.5 $ \AA,  $c_r = 20$ {\AA},  $\sigma^{(0)} = 0.12$ {\AA} and $c_\alpha^{(0)} = 0.3$ \AA$^2$/eV. We have run the EA multiple times with different hyperparameters values ($c_n$ from 4 {\AA} to 6.5 {\AA}, $c_r$ from 3 {\AA} to 50 {\AA}, $\sigma^{(0)}$ from 0.10 {\AA} to 0.2 {\AA} and $c_\alpha^{(0)}$ from 0 \AA$^2$/eV to 0.3 \AA$^2$/eV). In all cases the solution of the EA converged to the same defect configuration. 

\begin{figure} 
\subfloat{{\includegraphics[width=0.8\textwidth]{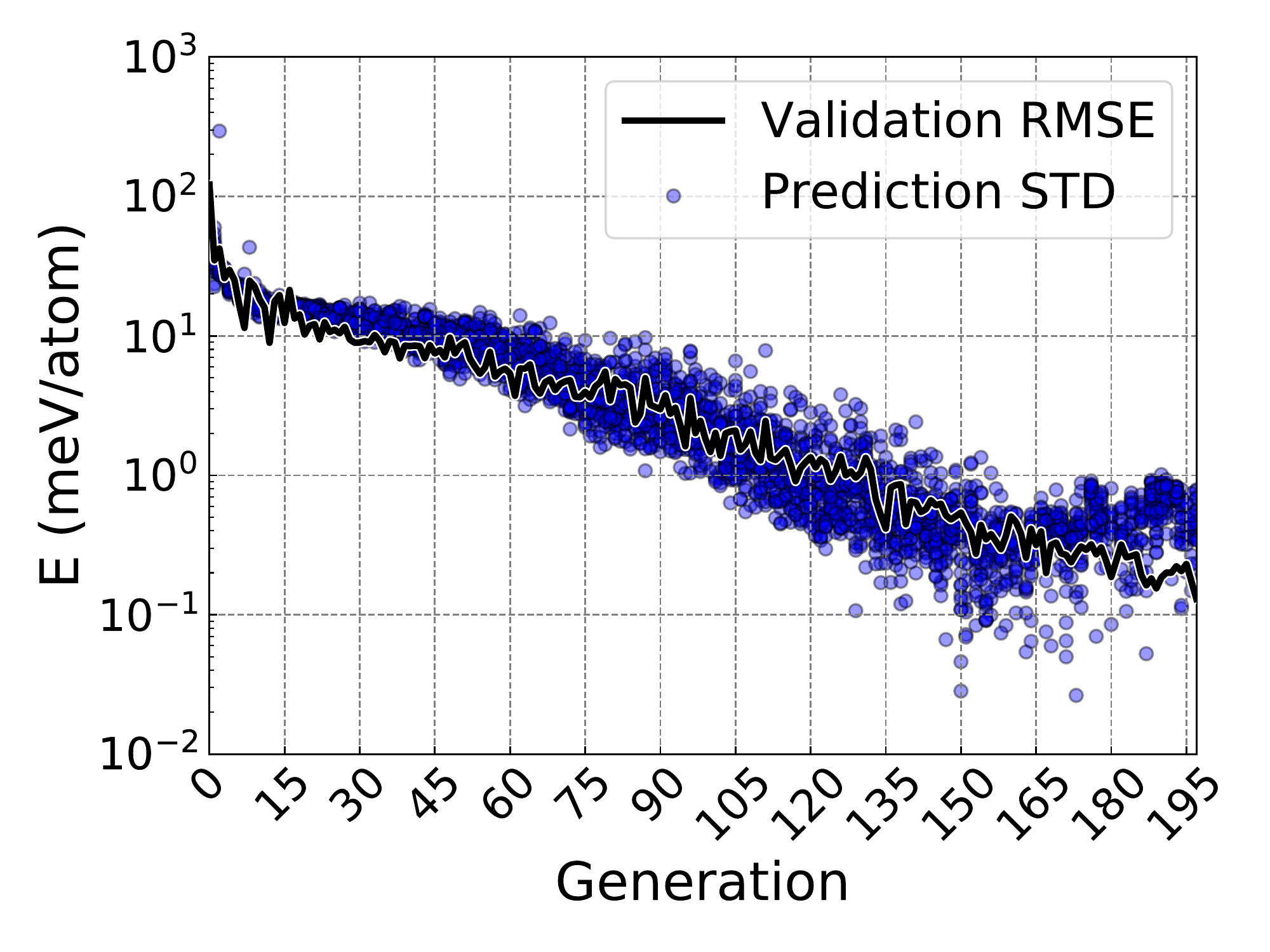}}} %
\caption{\label{fig:descriptors} Performance of the crystal fingerprint descriptors during the evolutionary process for   $\square_\text{O}$ in anatase TiO$_2$. Bold black lines represent the calculated root-mean-square-errors. Blue dots represents the values of the prediction standard error.}
\end{figure}
Employment of the metamodel does also lead to the same solution, if similar values for the hyperparameters are kept and the  threshold value for the prediction standard error is set below 10 meV/atom. As mentioned in the previous section, the main advantage of the metamodel consists in reducing the number of DFT calculations. As shows in Figure~\ref{fig:descriptors}, even with a complicated PES as the one produced by the DFT+$U$ method, the crystal fingerprint descriptors of Valle and Oganov allow to obtain accurate predictions of the DFT energy after few tens of generations.
In particular, the figure illustrates the validation root-mean-square-error (RMSE) and prediction standard errors as a function of the generation. The results at generation $g$ are obtained by training a GP regressor on the previous  generations. The energy values at generation $g$ are then used as the validation set where the RMSE is calculated. A PCA dimensionality-reduction step is performed and only the components necessary to preserve at least 95 \% of the variance are kept before training the GP. One can see that a prediction RMSE and standard error within 10 meV/atom are readily achieved after a couple tens of generations. As a matter of fact, it is in the nature of the EA that the more the evolution proceeds, the more similar structures are generated. Hence, the prediction RMSE  will ultimately decrease, as can be seen from equation~\eqref{eq:posterior_mean}, which shows that the prediction of a GP is a linear smoothing of the training set targets.

\subsection{\label{subsec:Results} Results}

\begin{figure}[t]
\centering
\subfloat{{\includegraphics[width=.8\textwidth]{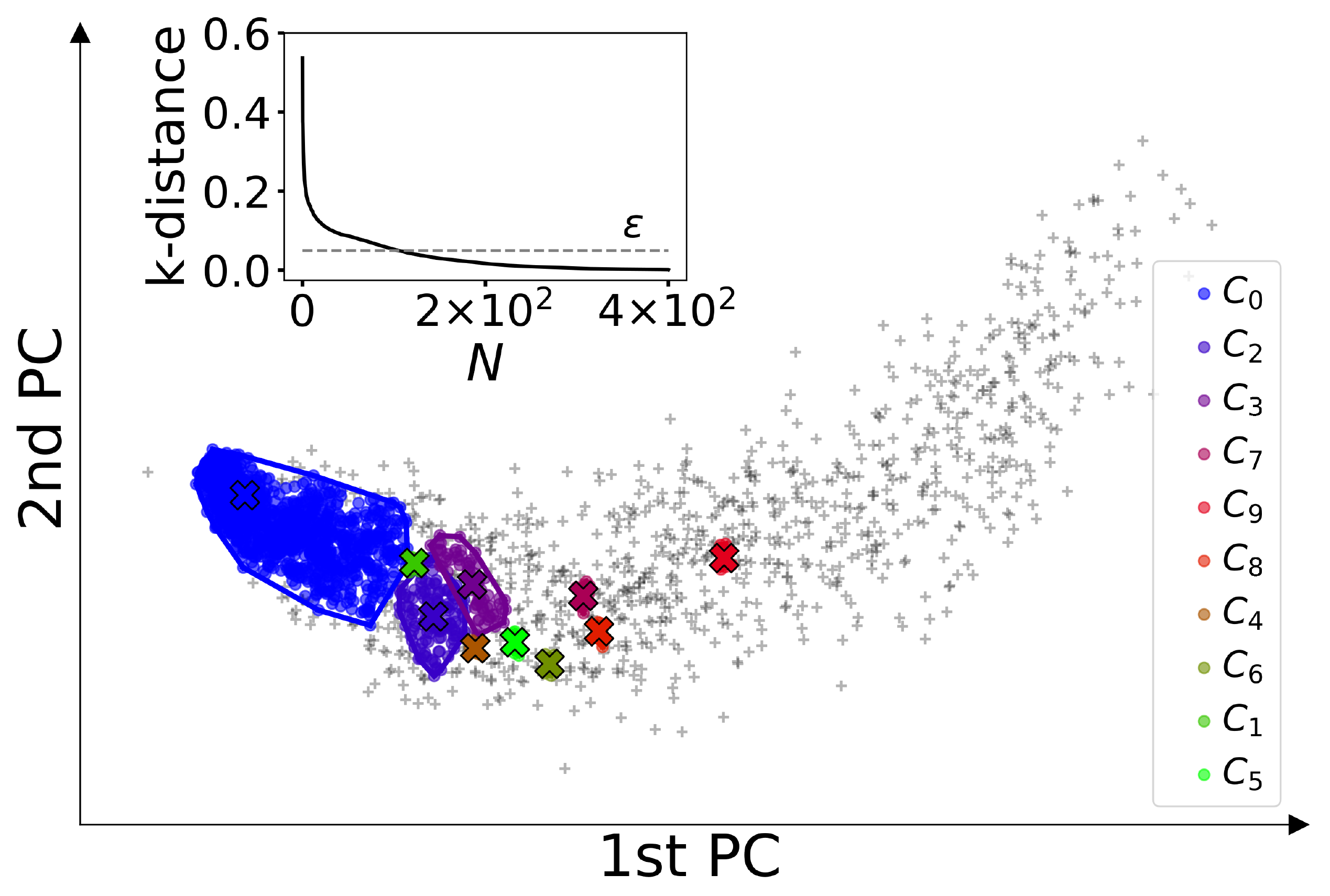}}} \\
\caption{\label{fig:tio2clusters} The clusters found among the visited structures of an EA run for the $\square_\text{O}$ in anatase TiO$_2$. Cluster centroids are shown with crosses. The inset shows the $k$-distance plot used to find the clustering parameter $\varepsilon$. Cluster names  are ordered in the legend according to their size. }
\end{figure}
With all the parameters described in the previous section, we found the EA to converge always to a configuration which is neither the simple nor the split vacancy one. This configuration resembles somewhat the simple vacancy, but has lower symmetry and considerable lower energy: its formation energy is only around 30 meV larger than that of the split vacancy. We named this configuration as the "pseudo-symmetric" vacancy.

\begin{figure}
\centering
\includegraphics[width=0.5\textwidth]{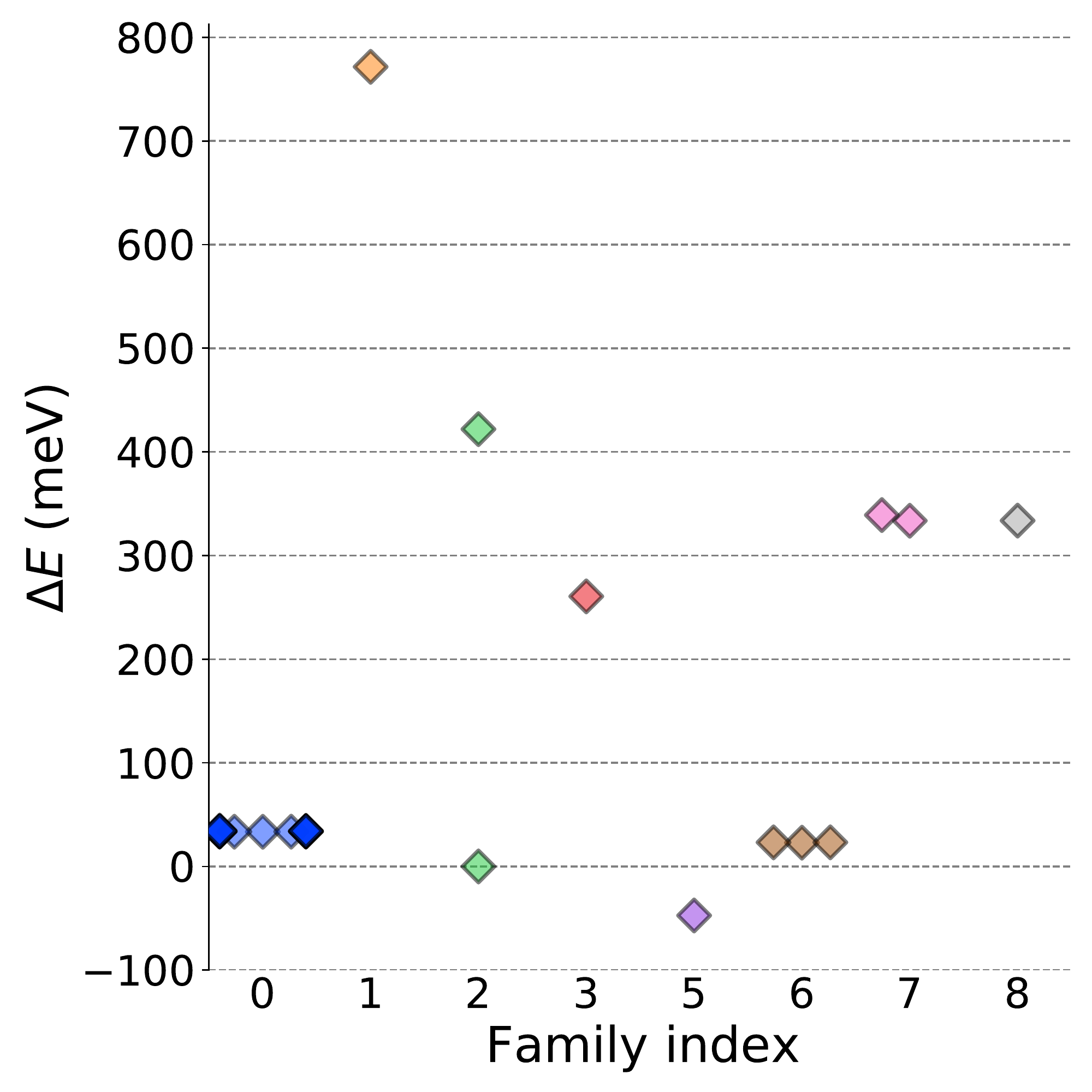}
\caption{\label{fig:tio2clusters2} Distribution of the minima found at the DFT+$U$ level for $\square_\mathrm{O}$ in TiO$_2$ anatase.}
\end{figure}
In order to look at other explored regions of the PES, we again represented the structures visited by the EA using the crystal fingerprints and plot the two principle components as a scatter plot in Figure~\ref{fig:tio2clusters}. We obtain an analogous cloud as that of Figure~\ref{fig:si_clusters}. Also in this case the first principal components strongly anticorrelates with the generation number so that the later generations are found to the left.
We employed the clustering approach described in section~\ref{subsec:CLUS} collecting all the structures visited by a run of the EA. We set the \texttt{DBSCAN} parameter $minPts$ to 8 and found the value of $\varepsilon$ by analyzing the 8-distance plot shown in the inset of Figure~\ref{fig:tio2clusters}. Several clusters were found, with  $C_0$ being the largest one since it contains individuals generated in the last generations, where the structures are similar to one another. Nine other clusters were found. As in Section \ref{subsec:ffcd}, their names reflects the magnitude of the Euclidean  distance from $C_0$. The behavior of relative average KL divergence and Bhattacharyya distance per cluster is similar to the one described by Figure~\ref{fig:si_clusters2}, but this time both quantities increase monotonically with the cluster's centroid distance from $C_0$.  In particular, both quantities also show a steep increase in magnitude already from $C_1$, illustrating that the algorithm is indeed able to separate the converged structure from others. According to these observations, we took the structures in $C_9$, $C_6$, which are relatively small clusters with large values of the relative $\langle D(C) \rangle$ and we also included the structures in  $C_5$, since the cluster is very small. The overall number of structures in these clusters is the following: ten for $C_9$, seven for $C_6$ and six for $C_5$. We relaxed all of them using a gradient-based optimization algorithm. Multiple new minima were found, including a configuration with energy lower than that of the split vacancy by around 50 meV, but with a closely related structure. As a matter of fact, most of the new local minima show a structure either similar to that of the pseudo-symmetric vacancy or to the split vacancy, but have distinguishable different levels of lattice distortions. Adding to these structures those of  the known symmetric and split vacancy configurations, we obtain a total of 25 structures, whose distribution in formation energy, with respect to that of the split vacancy, is shown in  Figure~\ref{fig:tio2clusters2}. The presence of several low-energy minima indicates that the PES described by the DFT+$U$ method is  artificially complex for this system. As the next step, we grouped all these 25 structures into nine families, according to the values of their first two principal components; specifically, the family with index 0, includes the pseudo-symmetric configuration, that with index 1 consists of the symmetric vacancy and that with index 2 includes the split vacancy. Other indices are assigned to families that do not comprise these configurations. From Figure~\ref{fig:tio2clusters2}, one can see that in general each family comprises structures with very similar formation energies. However, some exceptions appear: structures belonging to the same family might have a notably different energy and structures with very similar energy might belong to different families.  We then selected a representative for each of the nine families and we relaxed its structure employing the HSE15 functional. Among the structures calculated at the HSE15 level (exluding the known symmetric configuration), all except two (the representatives of family  3 and 6, both belonging to $C_9$) relaxed to the same split-vacancy configuration reported in reference \onlinecite{Arrigoni-2020} (including the various pseudo-symmetric vacancy configurations), showing that many of the minima appearing on the DFT+$U$ PES are indeed spurious and disappear when a more accurate functional is used. The remaining two structures relaxed to a configuration characterized by large displacements of the two oxygen atoms close to $\square_\text{O}$, which is neither the simple nor the split vacancy. We name this structure the "delocalized vacancy" configuration, as the extra electrons are delocalized in conduction band states, as shown in Figure~\ref{fig:geometries}. Notably at this level of accuracy, this configuration represents the ground state, with an energy of around 0.32 eV lower than that of the split vacancy.  

The existence of this configuration completes the atomistic description of the neutral $\square_\text{O}$ in TiO$_2$ anatase. In fact, considering the electronic structure of this defect, three different scenarios are possible: in the simple vacancy the two extra electrons are localized in the vacancy  site, in the split vacancy one electron is excited to a delocalized conduction-band state, while the other is strongly localized on a neighboring Ti atom, reducing it to the +3 charge state and  inducing strong lattice distortions. Finally, in the delocalized vacancy, both electrons are delocalized in the conduction band. 

\section{\label{sec:Conclusions} Conclusions}
In this work we have proposed an approach based on a variation of the CMA-ES algorithm optimized for finding low-energy structures of point defects in solids. We have shown that the algorithm is generally able to find the ground state defect configuration for a set of defects showing different electronic and structural properties. Using this algorithm, we were able to find known low-energy defective structures and new ones as well. The computational costs required by the algorithm are affordable even at the first-principles level, and can be further  reduced by employing a Gaussian process regressor which is trained on-the-flight during the evolutionary process and is devised to guarantee a minimum accuracy of the approximated PES. 

The application of our proposed approach also demonstrates how the DFT+$U$ leads to an overly-complex PES, with several spurious local minima, for the neutral  $\square_\mathrm{O}$ in TiO$_2$ anatase,  making the convergence of the EA to the global minimum difficult. However, the application of the unsupervised clustering post-processing step allows for an efficient selection of interesting structures, which can then be short-listed an studied with more accurate functionals. Employing this procedure, we were able to discover a new low-energy configuration for the oxygen vacancy in TiO$_2$ anatase, characterized by the presence of two electrons delocalized in the conduction band, which completes the atomistic description of this well-studied defect. 

A dataset containing the structures described in this work, including those generated during the evolutionary processes, is available on Zenodo, reference \onlinecite{data_zenodo}.

\begin{acknowledgments}
The authors acknowledge support from the Austrian Science Funds (FWF) under project CODIS (FWF-I-3576-N36). Part of the calculations were performed on the Vienna Scientific Cluster under the project 1523306: CODIS.
\end{acknowledgments}

%
\end{document}